\documentclass[12pt]{oxarticle}
\pdfoutput=1

\usepackage{graphicx, float, array, xspace, amscd, amsmath, amsthm, amssymb, latexsym, relsize, bbm}
\usepackage[a4paper, left=1.7in, right=0.3in, top=0.8in, bottom=1.2in]{geometry}
\usepackage[bbgreekl]{mathbbol}
\usepackage{amsfonts, tikz-cd}
\usepackage[T1]{fontenc}
\PassOptionsToPackage{linecolor=blue,backgroundcolor=blue!25,bordercolor=blue,textsize=scriptsize}{todonotes}
\usepackage{todonotes}
\usepackage{tikz}
\usetikzlibrary{backgrounds}
\usepackage[framemethod=tikz]{mdframed}
\AtBeginEnvironment{mdframed}{%
\tikzset{every picture/.style={}}%
}
\mdfsetup{roundcorner=.5ex}
{\begin{mdframed}[backgroundcolor=white]}%
{\end{mdframed}}

\DeclareSymbolFontAlphabet{\mathbb}{AMSb}
\DeclareSymbolFontAlphabet{\mathbbl}{bbold}

\usepackage[sort&compress, comma, square, numbers]{natbib}
\usepackage[all,cmtip]{xy}
\usepackage{color}
\definecolor{MyDarkBlue}{rgb}{0.15,0.25,0.45}
\usepackage[linktocpage=true]{hyperref}
\hypersetup{colorlinks=true, pageanchor=false, citecolor=blue, linkcolor=blue, urlcolor=blue, pdfauthor={}, pdftitle={},breaklinks=true}

\let\SS=\S 

\newcommand{\sh}{{\sharp}}

\newcommand{\Ric}{{\rm Ric}}
\def\Dbar{{\overline{D}}}

\newcommand{\ol}[1]{{\overline{#1}}}

\newcommand{\LC}{\text{\tiny LC}}
\newcommand{\CH}{\text{\tiny CH}}

\renewcommand{\sb}{{\overline{\sigma}}}

\newcommand{\kb}{{\overline{ \kappa}}}
\newcommand{\rb}{{\overline{ r}}}

\newcommand{\Ob}{{\overline{ \Omega}}}

\newcommand{\w}{{\,\wedge\,}}

\newcommand{\wh}{\widehat}
\newcommand{\ccZh}{{\,\wh{\ccZ}\,}}

\newcommand{\fD}{{\mathfrak{D}}}
\newcommand{\fDb}{{\overline\fD}}

\newcommand{\half}{\frac{1}{2}}
\newcommand{\qrt}{\frac{1}{4}}

\def\CS{{\text{CS}}}

\newcommand{\bb}{{\overline\beta}}

\renewcommand{\a}{\alpha}

\newcommand{\G}{\Gamma}
\renewcommand{\d}{\delta}\newcommand{\D}{\Delta}

\renewcommand{\k}{\kappa}
\renewcommand{\l}{\lambda}\renewcommand{\L}{\Lambda}
\newcommand{\m}{\mu}
\newcommand{\n}{\nu}

\renewcommand{\r}{\rho}
\newcommand{\s}{\sigma}\renewcommand{\S}{\Sigma}
\renewcommand{\t}{\tau}

\renewcommand{\o}{\omega}\renewcommand{\O}{\Omega}

\DeclareFontFamily{OT1}{pzc}{}
\DeclareFontShape{OT1}{pzc}{m}{it}{<-> s * [1.200] pzcmi7t}{}
\DeclareMathAlphabet{\mathpzc}{OT1}{pzc}{m}{it}

\newcommand{\cB}{\mathcal{B}}

\newcommand{\cE}{\mathcal{E}}
\newcommand{\cF}{\mathcal{F}}
\newcommand{\cG}{\mathcal{G}}
\newcommand{\cH}{\mathcal{H}}

\newcommand{\cK}{\mathcal{K}}
\newcommand{\cL}{\mathcal{L}}

\newcommand{\cO}{\mathcal{O}}
\newcommand{\ccP}{\mathpzc P}
\newcommand{\cQ}{\mathcal{Q}}

\newcommand{\cX}{\mathcal{X}}
\newcommand{\ccY}{\mathpzc Y}
\newcommand{\ccZ}{\mathpzc Z}

\newcommand{\ccZb}{{\overline \ccZ}}

\newcommand{\ccPb}{{\overline \ccP}}

\DeclareFontFamily{U}{bbold}{}
\DeclareFontShape{U}{bbold}{m}{n}
 {  <-5.5> s*[1.05] bbold5
    <5.5-6.5> s*[1.05] bbold6
    <6.5-7.5> s*[1.05] bbold7
    <7.5-8.5> s*[1.05] bbold8
    <8.5-9.5> s*[1.05] bbold9
    <9.5-11.5> s*[1.05] bbold10
    <11.5-16> s*[1.05] bbold12
    <16-> s*[1.05] bbold17
 }{}

\newcommand{\IR}{\mathbbl{R}}

\font\eightrm=cmr8 at 8pt
\font\csc=cmcsc10

\newcommand{\beq}{\begin{equation}}
\newcommand{\eeq}{\end{equation}}
\newcommand{\beqnn}{\begin{equation*}}
\newcommand{\eeqnn}{\end{equation*}}
\newcommand{\bea}{\begin{eqnarray}}
\newcommand{\eea}{\end{eqnarray}}
\newcommand{\bean}{\begin{eqnarray*}}
\newcommand{\eean}{\end{eqnarray*}}

\newcommand{\sref}[1]{\SS\ref{#1}}

\newcommand{\ee}{\text{e}}
\newcommand{\ii}{\text{i}}

\newcommand{\place}[3]{\vbox to0pt{\kern-\parskip\kern-7pt
                             \kern-#2truein\hbox{\kern#1truein #3}
                             \vss}\nointerlineskip}

\DeclareFontFamily{U}{wncy}{}
\DeclareFontShape{U}{wncy}{m}{n}{<->wncyr10}{}
\DeclareSymbolFont{mcy}{U}{wncy}{m}{n}
\DeclareMathSymbol{\sha}{\mathord}{mcy}{"58}

\newcommand{\del}{{\partial}}
\newcommand{\delb}{{\overline{\partial}}}

\newcommand{\jb}{{\overline\jmath}}
\newcommand{\lb}{{\overline\lambda}}
\newcommand{\nb}{{\overline\n}}
\newcommand{\mb}{{\overline\m}}

\newcommand{\Db}{{\overline \Delta}}

\renewcommand{\aa}{\mathfrak{a}}

\newcommand{\EndE}{{\text{End}\,E}}

\newcommand{\dd}{{\text{d}}}

\newcommand{\K}{K\"ahler\xspace}

\renewcommand{\H}{\text{H}}

\def\ker{{\rm ker ~}}

\newcommand{\vol}{\dd^6 x \sqrt{g}\, }
\newcommand{\volh}{\dd^6 x \sqrt{\wh g}\, }
\newcommand{\tr}{\text{tr}\hskip2pt}

\newcommand{\tb}{{\overline{\tau}}}

\newcommand{\ap}{{\a^{\backprime}\,}}
\renewcommand{\sb}{{\overline{\sigma}}}

\renewcommand{\rb}{{\overline{\rho}}}
\renewcommand{\=}{\;=\;}

\makeatletter
\g@addto@macro\bfseries{\boldmath}
\makeatother

\newcommand{\citeM}{\cite{Candelas:2016usb}\xspace}

\newcommand{\citeUG}{\cite{Candelas:2018lib, McOrist:2019mxh}\xspace}

\newcommand{\citeAP}{\cite{McOrist:2025zwf}\xspace}
\newcommand{\citeUGAP}{\cite{McOrist:2026boe}\xspace}

\newcommand{\citeMP}{\cite{McOrist:2025zwf}\xspace}
\newcommand{\citeDecoupling}{\cite{Chisamanga:2024xbm}\xspace}

\renewcommand{\baselinestretch}{1.1}
\numberwithin{equation}{section}
\proofmodefalse
\begin{document}
\pagestyle{empty}
\ifproofmode\underline{\underline{\Large Working notes. Not for circulation.}}\else{}\fi

\begin{center}
\null\vskip0.2in
{\Huge Heterotic moduli, the double extension and the $\alpha'{\,}^2$ metric  \\[0.5in]}

{\csc Jock McOrist and Qianhe Yin\\[0.2in]}

{\it Department of Mathematics\hphantom{$^2$}\\ School of Science and Technology\\ University of New England\\ Armidale, 2351, Australia\\[3ex]   }
\vspace{1cm}
{\bf Abstract\\[-8pt]}
\end{center}

We compute the heterotic moduli-space metric through $\ap^2$ for backgrounds admitting a smooth $\ap\to0$ limit.  The \K potential is unchanged when written in an appropriate field basis suggested by the ten-dimensional supersymmetry equations, while the metric receives $\ap^2$ corrections.  We discuss this and clarify the  roles of the extension bundle,  F-terms, the D-terms and the string-derived moduli space metric.

\vskip150pt

\newgeometry{left=1.5in, right=0.5in, top=0.75in, bottom=0.8in}
%
\newpage

%
%
\setcounter{page}{1}
\hypersetup{pageanchor=true}
\pagestyle{plain}
\renewcommand{\baselinestretch}{1.3}
\null\vskip-10pt

\section{Introduction}

The moduli of heterotic compactifications to four dimensions preserving $N=1$ supersymmetry are usually described by three pieces of data: a complexified hermitian deformation $\ccZ$, a bundle deformation $\aa$ and a complex-structure deformation $\D$. String-scale corrections are organised in powers of $\ap$. At zeroth order in $\ap$, the compactification reduces to one on a Calabi--Yau manifold: the bundle sector decouples, as do $\ccZ$ and $\D$. Although $\ap=0$ provides a useful formal limit, it is not itself a finite-tension string theory; in particular, the Green--Schwarz anomaly-cancellation mechanism enters at non-zero $\ap$. We therefore consider backgrounds admitting a smooth $\ap\to0$ limit while keeping $\ap$ small and non-zero.

At first order in $\ap$, the supersymmetry conditions on the internal geometry $X$, the bundle $E\to X$ and the three-form $H$ can be expressed as F-terms and D-terms together with the anomaly-cancellation condition. For these data, the ten-dimensional supersymmetry conditions of \cite{Bergshoeff:1989de,Bergshoeff:1988nn} take the compact geometric form derived by Hull and Strominger \cite{Hull:1986kz,Strominger:1986uh}. Crucially, within this formulation the tangent-bundle connection is not a free choice: supersymmetry fixes it to the Hull connection $\Theta^\H$ appearing in the Bergshoeff--de Roo action. The Bergshoeff--de Roo action is supported by independent comparisons with string scattering amplitudes and worldsheet beta functions. These agree with the action through order $\ap$ \cite{Cai:1986sa,Gross:1986mw,Hull:1986xn,Ross:1986ra,Hull:1987pc,Metsaev:1987zx}, with partial amplitude results and beta-function checks extending to order $\ap^2$ \cite{Metsaev:1986yb,Foakes:1988wy,Foakes:1987bn,Ellwanger:1988cc,Ellwanger:1987zx,Chemissany:2007he}.

Within this framework, the F-term constraints couple the moduli through a generalisation of the Atiyah map \cite{Atiyah:1955} and the trivialisation of $H$ \cite{delaOssa:2014cia,Anderson:2014xha}, while the D-terms select representatives of the resulting infinitesimal deformation classes \cite{McOrist:2021dnd}. Dimensional reduction then yields a metric on these physical representatives in the effective four-dimensional theory \citeM.

The purpose of this paper is to compute that metric through $\ap^2$, assuming that the heterotic background has a smooth $\ap\to0$ limit. It was proved in \citeAP that, for such backgrounds,
$$
H=\cO(\ap)~.
$$
The main result of \citeMP was that the flux--torsion, holomorphic-volume and conformally balanced equations retain the form derived in \cite{Hull:1986kz,Strominger:1986uh} through $\ap^2$ when written in a redefined field basis $(\widehat g,\widehat\o,\widehat\Phi)$, which we call the hatted basis and describe below. This field redefinition packages the $\ap^2$ corrections rather than removing them. Even in this basis, the graviton equation contains an explicit $\ap^2$ correction, although no separate bosonic term at this order appears in the Bergshoeff--de Roo action.

The tangent-bundle connection appearing in the action and Bianchi identity remains composite, with supersymmetry fixing it to be the Hull connection $\Theta^\H$. This connection is not a holomorphic instanton; its failure to be an instanton is essential for the supersymmetry integrability conditions to reproduce precisely the $\ap^2$ correction to the graviton equation \citeMP. Since $\Theta^\H$ is composite, its deformations are determined by  $\ccZ$, $\aa$ and $\D$. Their contribution produces $\ap^2$ corrections to the metric, so the resulting inner product is not simply the canonical $L^2$ norm.

We show that the functional form of the \K potential derived in \citeM is unchanged through $\ap^2$ when written in the hatted basis:
\beq\label{eq:KahlerPot}
K \= \! - \log\left(\ii\int_X\Omega\w\Ob\right) - \log\left(\frac{4}{3}\int_X\wh\o^3\right) + \cO(\ap^3)~.
\eeq
Its Hessian agrees with the metric obtained by dimensional reduction of the ten-dimensional action.  The metric is the large-radius supergravity approximation to the Zamolodchikov metric \cite{Zamolodchikov:1986gt}; worldsheet instantons and $g_s$ corrections lie outside the perturbative $\ap$ expansion considered here. Although the functional form of the \K potential is unchanged, the resulting metric contains non-trivial $\ap^2$ corrections. The deformation of the Hull connection is schematically
$$
\fD\Theta^\H \sim\nabla\D + \frac12\nabla\ccZ + \ap(\cdots)~,
$$
where the omitted terms involve $\aa$, $\ccZ$ and $\D$. Its norm produces the explicit $\ap$--corrections. 

A second goal is to distinguish three structures that enter the calculation. First, the hatted and unhatted variables are two perturbatively equivalent ten-dimensional field bases. Dimensional reduction is performed in the original Bergshoeff--de Roo variables, while the supersymmetry equations and the \K potential take their simplest form in the hatted variables. We therefore perform the reduction in the original variables and apply the field redefinition to the reduced metric. The quantities $H$, $\Theta^\H$, $R^\H$ and $\cE^\H$ retain their Bergshoeff--de Roo definitions throughout.

Second, the double extension \cite{McOrist:2021dnd,McOrist:2024zdz}
$$
0\longrightarrow T^{*(1,0)}X\longrightarrow\cQ\longrightarrow Q_1\longrightarrow0~, \qquad 0\longrightarrow{\rm End}\,E\longrightarrow Q_1\longrightarrow T^{1,0}X\longrightarrow0
$$
organises the F-term constraints. A choice of splitting writes a deformation as $(\wh\ccZ,\aa,\D)$, but need not give an orthogonal decomposition of the physical tangent space. The D-terms select physical representatives of the corresponding deformation classes.

Finally, the string-derived metric supplies an inner product on the finite-dimensional physical tangent space. A hermitian direction may have a physical representative
$$
E_\rho=(e_\rho,\aa_\rho,\D_\rho)~,
$$
with compensating bundle and complex-structure components. The hermitian directions and their complement can then be block-orthogonalised pointwise. This Gram--Schmidt projection is a pointwise change of frame on the physical tangent space; it is neither a ten-dimensional field redefinition nor, in general, a change of moduli coordinates.

The paper is organised as follows.  \sref{s:review} reviews the heterotic action, supersymmetry conditions, infinitesimal deformations and the F- and D-term representative equations. \sref{s:deformationHull} computes the deformation of the Hull connection to the order needed for the metric.  \sref{s:dimred} performs the dimensional reduction and isolates the explicit $\ap^2$ correction.  \sref{s:metric-final} assembles the corrected metric and gives its pointwise orthogonal decomposition on the physical tangent space. Finally, \sref{s:conclusion} summarises the result. Appendix~\ref{s:HullDetails} contains the Hull-connection calculation and the change between the hatted and unhatted $L^2$ pairings.

\section{Review of \texorpdfstring{$N=1$ $d=4$}{N=1 d=4} heterotic moduli physics}
\label{s:review}
This section fixes the conventions used below. We first review the ten-dimensional equations and the hatted variables natural for supersymmetry. We then introduce the unhatted deformation fields
$$
\ccY \= (\ccZ,\aa,\D)~,
$$
used in dimensional reduction, relate $\ccZ$ to $\wh\ccZ$, and state the F- and D-term representative equations. The labels ``complex-structure'', ``bundle'' and ``hermitian'' refer to the leading component in this splitting; the coupled supersymmetry conditions do not imply that only one structure varies.

\subsection{Heterotic action}

Up to the overall ten-dimensional Newton constant, the bosonic action in the field basis used below is
\beq
\label{eq:het-action}
S \= \frac{1}{2}\int \dd^{10}x\,\sqrt{g}\,\ee^{ - 2\Phi}\left\{ R - \frac{1}{2}|H|^2 + 4(\partial_m\Phi)^2 + \frac{\ap}{4}\left(\tr |F|^2 - \tr |R^\H|^2\right)\right\} + \cO(\ap^3)~.
\eeq
Here $F$ is the curvature of a connection $A$ on the gauge bundle $E$ and $R^\H$ is the curvature of the Hull connection
$$
\Theta^\H \= \Theta^\LC + \frac{1}{2}H~.
$$
The three-form $H$ satisfies the Green--Schwarz Bianchi identity
\beq
\label{eq:bianchi}
\dd H \= \frac{\ap}{4}\tr F\w F - \frac{\ap}{4}\tr R^\H\w R^\H~.
\eeq
Locally,
\beq
\label{eq:h-cs}
H \= \dd B + \frac{\ap}{4}\CS(A) - \frac{\ap}{4}\CS(\Theta^\H)~,
\eeq
where the gauge transformation of $B$ cancels the variation of the Chern--Simons forms. We assume that the compactification has a smooth $\ap\to0$ limit.  For supersymmetric backgrounds in this field basis this implies \citeMP
$$
H \= \cO(\ap)~, \qquad \nabla\Phi \= \cO(\ap)~.
$$

\subsection{Supersymmetry equations}
Let $X$ be the internal six-manifold in a compactification preserving $N=1$ supersymmetry. A globally defined nowhere-vanishing spinor on $X$ defines an $SU(3)$ structure consisting of an almost complex structure $J$, a hermitian form $\omega$ with compatible metric $g$, and a nowhere-vanishing $(3,0)$-form. The supersymmetry equations imply that $J$ is integrable and determine a holomorphic volume form $\Omega$. The $\ap^2$ corrections to this geometry were determined in \citeMP. To state the result, define
\beq
h \= \tr|F|^2 - \tr|R^\H|^2~, \qquad \cE^\H_{mn} \= \tr(F_{mp}F_n{}^p) - \tr(R^\H_{mp}\,R^{\H}{}_n{}^p)~,
\eeq
and the redefined metric, hermitian form and dilaton
\beq\label{eq:hatted-fields}
\widehat g_{\m\nb} \= g_{\m\nb} + \frac{\ap^2}{4}\cE^\H_{\m\nb}~, \qquad \widehat\o \= \o + \frac{\ii\ap^2}{4}\cE^\H_{\m\nb}\dd x^{\m\nb}~, \qquad \widehat\Phi \= \Phi + \frac{3\ap^2}{32}h~.
\eeq
Here $\o$ is the fundamental form determined by the Killing spinor, while $\wh\o$ is the fundamental form of $\wh g$ with respect to the same complex structure $J$. When it appears as a two-form, we use the shorthand $\cE^\H=\cE^\H_{\m\nb}\dd x^{\m\nb}$. We write $\star$ and $V=\int_X\dd^6x\sqrt g$ for the unhatted Hodge star and volume, and $\wh\star$ and $\wh V=\int_X\dd^6x\sqrt{\wh g}$ for their hatted counterparts. We work in constant dilaton gauge $\Phi = \Phi_0 + \cO(\ap^3)$, a consequence is $h=\cO(\ap)$ and hence $\wh\Phi=\phi_0 + \cO(\ap^3)$ and $\wh V=V + \cO(\ap^3)$ \citeMP.

Although the Hull--Strominger relations receive explicit $\ap^2$ corrections when written in the original fields, \citeMP showed that, up to terms of order $\ap^3$, they take the same form in the hatted variables:\footnote{Equality here means we drop terms $\ap^3$ or higher.}
\beq
\label{eq:susy-basic}
N_J \= 0~, \qquad F^{0,2} \= 0~, \qquad H \= \ii(\del - \delb)\wh\o~,
\eeq
together with
\beq
\label{eq:susy-moment-maps}
\dd(\|\Omega\|_{\wh g}\,\wh\o^2) \= 0~, \qquad F\w\wh\o^2 \= 0~.
\eeq
The first line is the F-term part of the system. The second line is the D-term part: the conformally balanced condition and the Hermitian Yang--Mills equation. These names are useful because the first line can be obtained from a holomorphic functional, while the second line has a moment-map interpretation. We only use these facts as a guide to terminology; in the calculation below the equations themselves are the input.

The hatted variables package some of the $\ap^2$ corrections rather than removing them. The dilaton is encoded by
\beq
\label{eq:omega-norm}
\log\|\Omega\|_{\wh g} \= - 2\wh \Phi + \cO(\ap^3)~.
\eeq
The Bianchi identity \eqref{eq:bianchi}, with the Hull connection, is imposed in addition to \eqref{eq:susy-basic}--\eqref{eq:susy-moment-maps}. In \citeMP these supersymmetry equations, together with the Bianchi identity, were shown to imply the ten-dimensional equations of motion through $\ap^2$, including the explicit correction to the graviton equation arising from the composite Hull connection.

Green--Schwarz anomaly cancellation fixes the tangent-bundle characteristic class, but not the connection used to represent it. In the Bergshoeff--de Roo field basis, the heterotic spectrum, ghost-freedom and local supersymmetry select the composite Hull connection $\Theta^\H$ \cite{Bergshoeff:1989de,Bergshoeff:1988nn}. A different connection can appear after a complete field redefinition, but the fundamental fields, $H$, the action and all composite quantities must then be redefined together; replacing $\Theta^\H$ only in the Chern--Simons and curvature terms is not such a redefinition. We retain the Bergshoeff--de Roo formulation, and the hatted variables above repackage the supersymmetry equations without replacing $\Theta^\H$ \cite{Melnikov:2014ywa}. Its deformation is therefore fixed by the physical fields rather than chosen independently.

The F-term equations admit a lift to the universal geometry over parameter space \cite{Candelas:2018lib,McOrist:2019mxh}. This construction remains valid through $\ap^2$ in the hatted basis, provided the universal Bianchi identity continues to use the Hull connection \citeUGAP. Infinitesimal solutions are represented by coupled deformations $(\wh\ccZ,\aa,\D)$ \cite{McOrist:2024glz}; their relation to the unhatted variables used in dimensional reduction is fixed below.

\subsection{Small deformations and notation}

Let $y^a$ be local coordinates on the space of solutions.  A small deformation of the background is written as
\beq
\label{eq:dg}
\d g_{mn} \= \d y^a\partial_a g_{mn}~, \qquad \d\omega_{mn} \= \d y^a\partial_a\omega_{mn}~.
\eeq
A complex-structure deformation is represented by
\beq
\label{eq:cs-deformation}
\dd x^\m \longmapsto \dd x^\m + \d y^a\D_{a\,\nb}{}^\m\,\dd x^{\nb}~.
\eeq
For the gauge bundle one differentiates covariantly along parameter space.  We write
\beq
\label{eq:DmodA}
\d A \= \d y^a\fD_a A~, \qquad \fD_a A \= \partial_a A - \dd_A\Lambda_a~,
\eeq
where $\Lambda_a$ is the parameter-space connection needed for gauge covariance.  

The $B$-field is best described through the gauge-invariant variation of $H$. Varying \eqref{eq:h-cs} gives
\beq
\label{eq:h-deformation}
\d H \= \d y^a\left( \dd\cB_a + \frac{\ap}{2}\tr(\fD_a A \w F) - \frac{\ap}{2}\tr(\fD_a\Theta^\H\w R^\H) \right)~.
\eeq
Here $\cB_a$ is the invariant two-form obtained by combining the variation of $B$ with the exact terms from the Chern--Simons variations. The notation used in the rest of the paper is
\beq
\label{eq:a-def}
\aa_a \= (\fD_a A)^{0,1}~.
\eeq
Thus $\aa_a$ is the bundle component of the infinitesimal deformation.  We use the antihermitian convention $A=A^{0,1} - (A^{0,1})^\dagger$ and denote the hermitian conjugate by $\ol\aa_a$.

The complexified hermitian deformation is
\beq
\label{eq:Za}
\ccZ_a \= \cB_a + \ii\partial_a\omega~.
\eeq
The three components of a deformation are therefore
$$
\ccZ_a~, \qquad \aa_a~, \qquad \D_a~.
$$

The moduli space is complex.  In holomorphic gauge we choose coordinates $y^a=(y^\a,y^{\bar\a})$ so that holomorphic tangent vectors are represented by
\beq
\label{eq:physical-dofs}
\begin{array}{rcl}
\D_{\a\nb}{}^{\m}\dd x^{\nb}~, && \hbox{complex-structure component}~,\\[3pt] \ccZ_\a^{1,1}~, && \hbox{complexified hermitian component}~,\\[3pt] \aa_\a~, && \hbox{bundle component}~.
\end{array}
\eeq
These labels refer to components in the field splitting; they do not imply that a physical modulus deforms only one field.  The antiholomorphic components are the conjugates of these fields and in this gauge
\beq
\label{eq:holoGauge}
\begin{aligned}
\aa_{\bar\a}& \= 0~, &\qquad \D_{\bar\a\nb}{}^{\m}& \= 0~,\\
\ccZ_{\bar\a}^{(1,1)}& \= 0~, &\ccZ_{\bar\a}^{(0,2)} \= \ol{\ccZ}_{\a}^{(0,2)} \= \ol{\ccZ}_{\bar\a}^{(2,0)}& \= 0~.
\end{aligned}
\eeq
We will usually suppress the holomorphic parameter index and write simply $\ccY=(\ccZ,\aa,\D)$.

In holomorphic gauge we have relations
\beq
\label{eq:SEfields}
\partial_\a g_{\bar\m\nb} \= 2\D_{\a\,(\bar\m\nb)}~, \qquad \cB_{\a\bar\m\nb} \= - \ii\partial_\a\omega_{\bar\m\nb} \= 2\D_{\a\,[\bar\m\nb]}~, \qquad \cB_{\a\m\nb} - \ii \partial_\a \o_{\m\nb} \= 0~.
\eeq
Later we will use that $\D_{\a\,[\mb\nb]} = \cO(\ap)$. 

We also need deformations of $\wh\o$:
\beq
\label{eq:hatted-Z}
\wh\ccZ_a \= \cB_a + \ii\partial_a\wh\o \= \ccZ_a - \frac{\ap^2}{4} \del_a \cE^\H~, \qquad {\ol {\wh \ccZ}}_a \= \ccZb_a + \frac{\ap^2}{4} \del_a \cE^\H~.
\eeq

The calculations in \sref{s:deformationHull} and \sref{s:dimred} use the original variables appearing in the Bergshoeff--de Roo action. In \sref{s:metric-final} we make the single change to the hatted variables used by the supersymmetry equations and the \K potential. Since $\wh\ccZ - \ccZ=\cO(\ap^2)$, replacing $\ccZ$ by $\wh\ccZ$ in a term with an explicit factor of $\ap$ changes that term only at order $\ap^3$. The $L^2$ norms require a separate check. Nonetheless, Appendix~\ref{s:hattedvsunhatted} shows that, when all quantities are transformed consistently, the hatted and unhatted $L^2$ pairings agree even through order $\ap^2$.

\subsection{F-terms and D-terms}

The `F-terms' are the linear deformations of \eqref{eq:susy-basic}. They are called such because the first-order deformations in $\ap$ come from a Gukov--Vafa--Witten-type superpotential \cite{LopesCardoso:2003dvb,delaOssa:2015maa,McOrist:2016cfl,Ashmore:2018ybe}. The linearisation of \eqref{eq:susy-basic} in the hatted basis is
\beq
\label{eq:fterms}
\begin{aligned}
\delb\D^\m + \cO(\ap^3) & \= 0~,\\
\delb_A\aa - F_\m\w\D^\m + \cO(\ap^3) & \= 0~,\\
\delb \wh \ccZ^{(1,1)} - 2\ii\D^\m(\del\wh\o)_\m + \frac{\ap}{2}\tr(F\aa) - \frac{\ap}{2}\tr( R^\H \fD\Theta^{\H\,0,1}) + \cO(\ap^3)& \= 0~,
\end{aligned}
\eeq
where indices are raised and lowered with $\wh g$. We use right contraction for suppressed form indices unless explicitly stated otherwise. Thus an $r$-form $X$ obeys
\beq
\label{eq:rightContr}
X_m=\frac{1}{(r - 1)!}X_{n_1\cdots n_{r - 1}m}\, \dd x^{n_1}\w\cdots\w\dd x^{n_{r - 1}} =( - 1)^{r - 1}\iota_{\partial_m}X~.
\eeq
In particular, for one-forms $A$ and $B$,
\beq
\label{eq:rightWedge}
(A\w B)_m=A\,B_m - A_m\,B~.
\eeq
Therefore $F_m=F_{nm}\dd x^n$ and, in complex coordinates, $F_\m= - F_{\m\nb}\dd x^{\nb}$; the same convention is used for $R^\H_m{}^{ab}$.

The second equation in \eqref{eq:fterms} says that a complex-structure component $\D$ can lift to a bundle deformation only if the Atiyah obstruction vanishes. The third says that the resulting pair generally requires a hermitian component $\wh \ccZ$. Thus a typical F-term construction has the schematic form
$$
\D\ \rightsquigarrow\ (\aa,\D)\ \rightsquigarrow\ (\wh \ccZ,\aa,\D)~,
$$
or, for a bundle deformation,
$$
\aa\ \rightsquigarrow\ (\wh \ccZ,\aa,0)~.
$$
These are F-term lifts; the squigle arrow represents embedding into extension bundles. As it stands only the F-terms are imposed; the D-terms will be imposed later, their job is to pick out representatives.

More precisely, the F-term deformation problem is organised by the double extension
\beq
\label{eq:Qbundle}
0\longrightarrow T^{*(1,0)}X\longrightarrow\cQ\longrightarrow Q_1\longrightarrow0~, \qquad 0\longrightarrow{\rm End}\,E\longrightarrow Q_1\longrightarrow T^{1,0}X\longrightarrow0~.
\eeq
In the remainder of this subsection we write $\ccY=(\wh\ccZ,\aa,\D)$. Since the extension and adjoint description used below are constructed only through order $\ap$, this agrees with the unhatted splitting at the relevant order. The three component equations in \eqref{eq:fterms} are then written compactly as
\beq
\label{eq:Dbar-ccY}
\Dbar\ccY \= 0~, \qquad \Dbar^2 \= 0\quad\hbox{through order $\ap$}~,
\eeq
where nilpotence follows from the Bianchi identity. A deformation class represented by a $Q_1$-valued $(0,1)$-form lifts to $\cQ$ precisely when the third equation in \eqref{eq:fterms} can be solved for $\wh\ccZ$. 

The final equation in \eqref{eq:fterms} also shows why the Hull connection appears in the problem.  Its variation is not an independent modulus.  It must be evaluated in terms of $\wh \ccZ$, $\aa$ and $\D$.  Through zeroth order in $\ap$,
\beq
\label{eq:theta-da-first}
\fD\Theta^{\H\,0,1}{}^\n{}_\m \= \nabla^-_\m\D{}^\n + \frac{1}{2}\nabla^{-\,\n}\wh \ccZ_{\m\bar\rho}\,\dd x^{\bar\rho} + \cO(\ap)~.
\eeq
Section~\ref{s:deformationHull} refines this formula by keeping the order-$\ap$ terms which contribute to the metric at order $\ap^2$.

The D-terms are the first variations of the Hermitian Yang--Mills and conformally balanced conditions. They choose representatives inside the F-term classes. Through order $\ap$, they are \cite{McOrist:2021dnd}:
\beq
\label{eq:dterms}
\begin{aligned}
\delb_A^\dagger\aa + \frac{1}{2}F^{\m\nb}\wh \ccZ_{\m\nb} + \cO(\ap^3) & \= 0~,\\[3pt]
\delb^\dagger\wh  \ccZ^{(1,1)}  + \cO(\ap^3)& \= 0~,\\[3pt]
\delb^\dagger\D^\n + \frac{1}{2}H^{\n\rho\bar\m}\wh \ccZ_{\rho\bar\m} - \frac{\ap}{4}\tr(F^{\n\bar\m}\aa_{\bar\m})  \qquad\qquad\qquad&\\[5pt]
  - \frac{\ap}{2}R^\n{}_{\rho}{}^{\s\bar\t}\nabla^-_\s\D_{\bar\t}{}^\rho + \frac{\ap}{4}R^{\n\rho\bar\s\bar\t}\nabla^-_\rho\wh \ccZ_{\bar\s\bar\t} + \cO(\ap^2) & \= 0~.
\end{aligned}
\eeq
Here $R$ is the Levi--Civita curvature. In terms with an $\ap$, changing between hatted and unhatted contractions affects only higher orders. The  D-terms are  complete through order $\ap$. The $\ap^2$ completion of the third equation, and hence a complete $\ap^2$ identification with the adjoint of an extended $\Dbar$ operator, is not constructed here.

Let $\ccY\in\Omega^{0,1}(\cQ)$ be any $\Dbar$-closed representative of a deformation class. Another representative of the same class is
$$
\ccY + \Dbar\Lambda~, \qquad \Lambda\in\Omega^{0,0}(\cQ)~.
$$
Imposing the D-term condition $\Dbar^\dagger(\ccY + \Dbar\Lambda)=0$ gives
\beq
\label{eq:Hodge}
\Dbar^\dagger\Dbar\Lambda \= \! - \Dbar^\dagger\ccY~.
\eeq
The complex on compact $X$ is elliptic \cite{McOrist:2024zdz,deLazari:2024zkg}, so the right hand side is orthogonal to $\ker\Dbar$ on degree-zero sections, so \eqref{eq:Hodge} has a unique solution on its orthogonal complement\footnote{The operator $\Dbar^\dag \Dbar$ has zero modes: degree zero sections $\k$ such that $\Dbar \k = 0$. For every such zero mode
$$
\langle\k , - \Dbar^\dag \ccY_F \rangle =\! - \langle \Dbar \k , \ccY_F \rangle = 0~,
$$
and so $\Dbar^\dag \ccY_F$ is orthogonal to all zero modes. Since $X$ is compact and the complex is elliptic, the operator can be inverted on the orthogonal complement of the zero modes. Hence, we can find a unique solution on the orthogonal complement
$$
\L =\! - \bigl( \Dbar^\dag \Dbar \bigr)^{ - 1} \Dbar^\dag \ccY_F~.
$$
}. Physically this says that every solution of the F-terms $\ccY$ can be adjusted by an exact deformation  $\Dbar \L$ so that it also satisfies the D-term condition $\Dbar^\dag (\ccY + \Dbar \L) = 0$. The combination of F-terms and D-terms imply $\ccY + \Dbar\Lambda$ is $\Dbar$-harmonic \cite{McOrist:2021dnd}. Here, $\Dbar$-harmonic refers only to the order in $\ap$ considered. The metric calculation does not require the higher-order form of $\Dbar$.

\section{Deformation of the Hull connection at \texorpdfstring{$\ap^2$}{alpha-prime squared}}
\label{s:deformationHull}
The Hull-connection norm enters the metric with an explicit factor of $\ap$, so we require $\fD\Theta^\H$ through order $\ap$ in the original Bergshoeff--de Roo variables. Since $\wh\ccZ - \ccZ=\cO(\ap^2)$, using $\wh\ccZ$ here would change the metric only at order $\ap^3$. We therefore use unhatted variables throughout this section.

\subsection{First order result}

We use $\G$ for coordinate connections and $\Theta$ for Lorentz-frame connections.  The coordinate connection is written
$$
\G_m{}^q{}_p~,
$$
where $m$ is the one-form index and ${}^q{}_p$ acts on vector indices.  The Hull connection is
$$
\G^\H_m{}^q{}_p \= \G^\LC_m{}^q{}_p + \frac{1}{2}H_m{}^q{}_p~.
$$
Thus
\beq
\label{eq:hull-def}
\d\G^\H_q{}^s{}_t \= \d\G^\LC_q{}^s{}_t + \frac{1}{2}\d H_q{}^s{}_t~.
\eeq
At leading order in $\ap$,
\beq
\label{eq:dHull0}
\begin{aligned}
\d\G^\LC_q{}^s{}_t & \= g^{sp}\left(\nabla_{[t}\d g_{p]q} + \frac{1}{2}\nabla_q\d g_{tp}\right)~,\\
\d H_q{}^s{}_t & \= - g^{sp}\d g_{pn}H_q{}^n{}_t + g^{sp}\left(\nabla_q\cB_{pt} + \nabla_p\cB_{tq} + \nabla_t\cB_{qp}\right) + \cO(\ap)~.
\end{aligned}
\eeq
It is useful to combine the metric and $B$-field variations into
\beq
\label{eq:Mdef-real}
M_p \= M_{qp}\dd x^q~, \qquad M_{qp} \= \frac{1}{2}(\d g_{qp} + \cB_{qp})~.
\eeq
Then the coordinate variation takes the compact form
\beq
\label{eq:first-hull-M}
\d\G^\H{}^s{}_t \= \dd_\nabla^\H M^s{}_t + \nabla_tM^s - \nabla^sM_t + \cO(\ap)~.
\eeq
The first term is covariant in coordinate indices but it is not part of the antisymmetric spin connection which appears in the action.  Indeed
\beq
\label{eq:ThetaGamma}
\Theta^\H_{mab} \= E_a{}^pE_b{}^q\G^\H_{mpq} - E_b{}^n\partial_m e_{an}~, \qquad \G^\H_{mpq} \= g_{pr}\G^\H_m{}^r{}_q~.
\eeq
On varying this relation, the $\dd_\nabla^\H M$ piece cancels against the vielbein variation.  The first-order Lorentz-frame deformation is therefore the metric-compatible antisymmetric part
\beq
\label{eq:dtheta-first}
\d\Theta^\H_{mab} \= E_a{}^pE_b{}^q\left(\nabla_{[q}\d g_{m|p]} + \nabla_{[q}\cB_{m|p]}\right) + \cO(\ap)~.
\eeq
Equivalently,
\beq
\label{eq:tr-dtheta-R-first}
\ap\tr(\d\Theta^\H\w R^\H) \= \ap R^{\H\,ts}\w\nabla_t(\d g + \cB)_{qs}\dd x^q + \cO(\ap^2)~.
\eeq
This is the term which enters the first variation of the Green--Schwarz correction.

\subsection{Second order result}
At order $\ap$, the variation of the frame factors multiplying $H$ must be retained along with the Chern--Simons variations. Using \eqref{eq:tr-dtheta-R-first}, the part of $\d H$ needed for the metric is
\beq
\label{eq:dHull1}
\begin{aligned}
\d H_q{}^s{}_t & \= g^{sm}\d g_{mn}g^{np}H_{qtp} + g^{sp}\left(\nabla_q\cB_{pt} + \nabla_p\cB_{tq} + \nabla_t\cB_{qp}\right)\\
&\quad + \frac{\ap}{2}g^{sp}\tr\left(\d A_qF_{pt} + \d A_pF_{tq} + \d A_tF_{qp}\right)\\
&\quad - \ap g^{sp}\left((\nabla_mM_{qn})R^\H{}_{pt}{}^{mn} + (\nabla_mM_{pn})R^\H{}_{tq}{}^{mn} + (\nabla_mM_{tn})R^\H{}_{qp}{}^{mn}\right)~.
\end{aligned}
\eeq
Combining this with the Levi--Civita variation, 
$$
\d\Theta^\H_{mab} = E_a{}^pE_b{}^q\Big(g_{pr}\d\G^\H_m{}^r{}_q - \nabla^\H_mM_{pq}\Big)~,
$$
 and choosing the Lorentz frame $\d e_{ap}=E_a{}^qM_{qp}$ gives
\beq
\label{eq:dtheta-second}
\begin{split}
 \d\Theta^\H_{ab} & \cong E_{a}{}^pE_{b}{}^q\Bigg[ \nabla^-_qM_{mp} - \nabla^-_pM_{mq} - H_{pq}{}^rM_{mr} + \frac{\ap}{4}\tr\left(\d A\,F_{pq} + \d A_qF_p - \d A_pF_q\right)\\[6pt]
&\qquad - \frac{\ap}{2}\left((\nabla_mM_n)R^\H{}_{pq}{}^{mn} + (\nabla_mM_{qn})R^\H{}_{p}{}^{mn} - (\nabla_mM_{pn})R^\H{}_{q}{}^{mn}\right)\Bigg]~,
\end{split}\raisetag{1cm}
\eeq
where $F_p=F_{qp}\dd x^q$ and $R^\H_q{}^{mn}=R^\H_{pq}{}^{mn}\dd x^p$ and  $\cong$ denotes up to $\ap^2$. The variation of the inverse vielbeins $E$ multiplying $H$, together with the conversion to $\nabla^-$, gives the displayed $HM$ term. The terms proportional to $\ap$ are the gauge and tangent-bundle Chern--Simons corrections. 

We now pass to complex coordinates.  From \eqref{eq:SEfields}, a holomorphic parameter-space variation has
\beq
\label{eq:Mcpx}
M_{\bar\m\nb} \= \d y^\a\D_{\a\,\bar\m\nb}~, \qquad M_{\bar\m\n} \= \d y^\a\frac{1}{2}\ccZ_{\a\,\bar\m\n}~,
\eeq
and $M_{\m p}=0$.  We split
$$
\d\Theta^\H \= \d y^\a\fD_\a\Theta^\H + \d y^{\bar\a}\fD_{\bar\a}\Theta^\H \= \fD\Theta^\H + \fDb\Theta^\H~,
$$
where the last equality suppresses the explicit $\d y$.  The holomorphic component of \eqref{eq:dtheta-second} is

\beq
\label{eq:dthcpx}
\begin{aligned}
\fD\Theta^{\H\,0,1}_{\nb\m}& \= \left(\ccP_{\m\nb\tb} + \ap C^F_{\m\nb\tb} + \ap C^R_{\m\nb\tb}\right)\dd x^\tb + \cO(\ap^2)~,\\[3pt]
\fD\Theta^{\H\,0,1}_{\m\n}& \= \! - \half\left(\nabla^-_\m\ccZ_\n - \nabla^-_\n\ccZ_\m + H_{\m\n}{}^\r\ccZ_\r\right) + \cO(\ap^2)~,\\[5pt]
\fD\Theta^{\H\,0,1}_{\mb\nb}& \= \! - \nabla^-_\mb\D_\nb + \nabla^-_\nb\D_\mb - H_{\mb\nb}{}^{\rb}\D_\rb + \cO(\ap^2)~.
\end{aligned}
\eeq
The components in the last line contribute at $\ap^3$ to the moduli space metric and so are not needed. The components $\ccP$, $C^F$ and $C^R$ are given in Appendix~\ref{s:Hullconnnorm}. The pure components are both order $\ap$ by the F-term and D-term equations.

\section{Dimensional reduction at \texorpdfstring{$\ap^2$}{alpha-prime squared}}
\label{s:dimred}

We now dimensionally reduce the ten-dimensional action in its original, unhatted Bergshoeff--de Roo variables. The result is converted to hatted variables in \sref{s:metric-final}; details are given in Appendix~\ref{s:HullDetails}. We follow the logic of \citeM.

The four-dimensional moduli-space metric is read off from the terms in the ten-dimensional action containing two spacetime derivatives of the moduli fields. We promote the parameters $y^a$ to spacetime fields $y^a(\cX)$ on $\IR^{3,1}$ and keep the massless $\cH$-singlet modes, where $\cH$ is the unbroken spacetime gauge group. For example, the standard embedding into one $E_8$ factor gives $\cH=E_6$.

The reduction ansatz is
\beq
\label{eq:kk-ansatz}
\begin{aligned}
\dd s^2 & \= \left(g_{ef} + \delta g_{ef}(\cX)\right)\dd\cX^e\otimes\dd\cX^f + \left(g_{mn}(x) + \delta g_{mn}(x,y(\cX))\right)\dd x^m\otimes\dd x^n~,\\
B & \= \delta B_{ef}(\cX)\dd\cX^e\w\dd\cX^f + \left(B_{mn}(x) + \delta B_{mn}(x,y(\cX))\right)\dd x^m\w\dd x^n~,\\
\Phi & \= \phi_0 + \varphi(\cX) + \phi(x,y(\cX))~.
\end{aligned}
\eeq
The internal coordinates are $x^m$ and the spacetime coordinates are $\cX^e$.

We first recall the terms already present at zeroth order in $\ap$.  After the Weyl rescaling to four-dimensional Einstein frame, the Einstein--Hilbert term gives
\beq
\label{eq:ricci-reduction}
\cL_g \= - \frac{1}{4V}\int_X\dd^6x\sqrt g\, (\partial_a g_{mn})(\partial_b g^{mn})\,\partial_e y^a\partial^e y^b~.
\eeq

The $H$ kinetic term gives
\beq
\label{eq:H-reduction}
\cL_H \= - \frac{1}{4V}\int_X\dd^6x\sqrt g\, \cB_{a\,mn}\cB_b{}^{mn}\,\partial_e y^a\partial^e y^b~.
\eeq
Using the holomorphic gauge conditions \eqref{eq:holoGauge} and the leading identifications \eqref{eq:SEfields}, their sum becomes
\beq
\label{eq:gH-reduction}
\cL_g + \cL_H \= \! - \frac{2}{V}\int_X\left( \D_\a{}^\m\star\ol\D_\bb{}^{\nb}g_{\m\nb} + \frac{1}{4}\ccZ_\a^{1,1}\star\ol\ccZ_\bb^{1,1} \right)\partial_e y^\a\partial^e y^{\bb}~.
\eeq
Thus the two universal kinetic terms give the familiar complex-structure and complexified hermitian inner products.

The Yang--Mills term contributes
\beq
\label{eq:gauge-reduction}
\cL_F \= \frac{\ap}{2V}\int_X\tr(\aa_a\star\ol\aa_b)\, \partial_e y^a\partial^e y^b~.
\eeq
In holomorphic gauge it pairs $\aa_\a$ with $\ol\aa_\bb$.  The curvature term for the Hull connection contributes
\beq
\label{eq:theta-reduction}
\cL_\Theta \= - \frac{\ap}{2V}\int_X\tr(\fD_\a\Theta^\H\star\fD_\bb\Theta^\H)\, \partial_e y^\a\partial^e y^{\bb}~.
\eeq
Putting the four pieces together defines the string derived moduli-space metric by
\beq
\label{eq:kinetic-final}
\cL_g + \cL_H + \cL_F + \cL_\Theta \= \! - 2g^{\sh}_{\a\bb}\,\partial_e y^\a\partial^e y^{\bb}~,
\eeq
with
\beq
\label{eq:moduli-metric}
\begin{aligned}
g^{\sh}_{\a\bb} \= \frac{1}{V}\int_X\Bigg(& \D_\a{}^\m\star\ol\D_\bb{}^{\nb}g_{\m\nb} + \frac{1}{4}\ccZ_\a^{1,1}\star\ol\ccZ_\bb^{1,1} - \frac{\ap}{4}\tr(\aa_\a\star\ol\aa_\bb)\\
&\qquad + \frac{\ap}{4}\tr(\fD_\a\Theta^\H\star\fD_\bb\Theta^\H)\Bigg) + \cO(\ap^3)~.
\end{aligned}
\eeq
{\bf Remark}: the minus sign in the gauge term is a convention: with antihermitian gauge fields the trace is negative definite on the Lie algebra. This convention differs from that of \citeUG but agrees with \citeAP, whose appendix discusses the sign. 

The first term can be written topologically \cite{Candelas:1990pi}
\beq
\frac{1}{V}\int_X \D_\a{}^\m\star\ol\D_\bb{}^{\nb}g_{\m\nb}  \= \! - \frac{\int_X \chi_\a \w \chi_\bb}{\int_X \Omega \w \ol \Omega}~, \qquad \chi_\a \= \D_\a{}^\m \O_\m~.
\eeq
This makes it clear this part of the metric is independent of the hermitian form. 

{\bf Remark}: the last term is not a metric on independent tangent-bundle deformations. In heterotic string theory $\fD\Theta^\H$ is composite and must be substituted in terms of $\ccZ$, $\aa$ and $\D$. The resulting metric is the large-radius approximation to the positive definite Zamolodchikov metric on the physical conformal manifold \cite{Zamolodchikov:1986gt,Kutasov:1988xb,Friedan:2012ky,Candelas:2016usb}.\footnote{To see this let $\phi_i(z)$ be a collection of real exactly marginal operators in the 2D unitary CFT. Their two-point function computes the Zamolodchikov metric for the moduli space of CFTs:
$$
\langle \phi_i(z) \phi_j(0)\rangle = \frac{g_{ij}}{|z|^4} + \cdots~,
$$
and for any vector $v^i$, $v^i v^j g_{ij} = |z|^4 \langle \phi_v(z) \phi_v(0) \rangle = \langle v | v \rangle \ge 0$, where $|v\rangle = \phi_v(0)|0\rangle$. In a unitary theory the Hilbert-space norm is positive by reflection positivity, or equivalently by radial quantisation. The Zamolodchikov metric is therefore positive definite after redundant directions have been removed. }  This differs from indefinite pairings on separate moduli problems in which the tangent-bundle connection is varied independently \cite{Garcia-Fernandez:2020awc,Garcia-Fernandez:2024hmf}; such theories have negative definite kinetic terms and so the theory has no ground state.

\subsection{Curvature corrections to the metric}
\label{s:HullCalc}
The mixed component in \eqref{eq:dthcpx} determines the norm through order $\ap^2$. Appendix~\ref{s:Hullconnnorm} evaluates it by integration by parts, using the F-term, D-term and constant-dilaton identities. It turns out to be convenient to express quantities in terms of hermitian connections: Chern on $\ccZ$ and a product connection of Bismut and Chern on $\D$ as described in the appendix. The end result is
\beq
\label{eq:HullMetric}
\begin{aligned}
&\frac{\ap}{4V}\int_X\tr\!\left(\fD_\a\Theta^\H\star\fD_\bb\Theta^\H\right)\\[3pt]
&\qquad\cong - \frac{\ap}{2V}\int_X\vol\,\left(\D_{\a\,\mb\nb}\ol\D_{\bb\,\s\r} + \qrt\ccZ_{\a\,\r\mb}\ccZb_{\bb\,\s\nb}\right)R^{\H\,\s\mb\r\nb} + \cO(\ap^3)~.
\end{aligned}
\eeq
Here and below, $\cong$ denotes equality modulo total derivatives, F- and D-terms and terms of order $\ap^3$ in the metric. At first order, $R^\H$ may be replaced by the Levi--Civita curvature, recovering the result of \citeM.

\section{The moduli space metric to \texorpdfstring{$\ap^2$}{alpha-prime squared}}
\label{s:metric-final}
The dimensional reduction in \sref{s:dimred} gives the metric in the original Bergshoeff--de Roo splitting $(\ccZ,\aa,\D)$. We now express the final result in the hatted splitting natural for the $\ap^2$-corrected supersymmetry equations and the \K potential:
\beq
\label{eq:hattedSplitting}
\ccY \= (\wh\ccZ,\aa,\D) \in \Omega^{0,1}\!\left(T^{*(1,0)}X\oplus{\rm End}\,E\oplus T^{1,0}X\right)~.
\eeq
The relation between $\wh\ccZ$ and $\ccZ$ is \eqref{eq:hatted-Z}. Appendix~\ref{s:hattedvsunhatted} shows that the two unsuppressed $L^2$ pairings are unchanged through order $\ap^2$ once the field redefinition and the change of Hodge star are both included. We therefore write these terms using $\wh g$, $\wh\star$ and $\wh V$. In terms carrying an explicit factor of $\ap$, replacing $\ccZ$ by $\wh\ccZ$ changes the metric only at order $\ap^3$. For consistency we keep the hat on $\ccZ$ throughout this section, while $R^\H$ and the contractions  in the explicitly suppressed terms retain their original Bergshoeff--de Roo definitions.

\subsection{Metric and \K potential}
\label{s:components}

Substituting \eqref{eq:HullMetric} into \eqref{eq:moduli-metric} gives the metric through order $\ap^2$ in the original variables. We now write it in the hatted basis:
\beq
\label{eq:metric-final-ap2}
\begin{aligned}
&g^{\sh}_{\a\bb} \cong \frac{1}{\wh V}\int_X\Bigg\{\D_\a{}^\m~\wh\star~\ol\D_{\bb}{}^{\nb}~\wh g_{\m\nb} + \frac14\wh\ccZ_\a^{1,1}~\wh\star~\ol{\wh\ccZ}_{\bb}^{1,1} - \frac{\ap}{4}\tr\!\left(\aa_\a\star\ol\aa_\bb\right)\Bigg\}\\[5pt]
&\qquad\qquad - \frac{\ap}{2V}\int_X\vol\left(\D_{\a\,\mb\nb}\ol\D_{\bb\,\s\r} + \qrt\wh\ccZ_{\a\,\r\mb}\ol{\wh\ccZ}_{\bb\,\s\nb}\right)R^{\H\,\s\mb\r\nb} + \cO(\ap^3)~.
\end{aligned}
\eeq
Here $R^\H$ is the Hull curvature. The first line gives the standard $L^2$ pairings, which would have been the natural mathematical ansatz for a metric. In the second line we see how string theory modifies that intuition: it induces curvature couplings, or interaction terms at order $\ap$ and $\ap^2$. At order $\ap$, one may set $R^\H=R$, recovering the result of \citeUG. 

Appendix~\sref{s:KahlerPot} derives the same metric from \eqref{eq:KahlerPot}. In the hatted basis the calculation has the same form as in \citeM. Thus \eqref{eq:KahlerPot} is the \K potential through order $\ap^2$.

\subsection{Double extension review}

In the splitting \eqref{eq:hattedSplitting}, write $\ccY=(\ccZh,q)$, where $q=(\aa,\D)$. The operator on $\cQ$ is
\beq
\label{eq:Dbar-closed}
\Dbar \=
\begin{pmatrix}
\cL_\ccZh & \cE\\ 0 & \Dbar_1
\end{pmatrix}
~, \qquad \Dbar^2 \= 0 \quad\hbox{through order $\ap$}~.
\eeq
On $Q_1$,
\beq
\label{eq:Dbar1}
\Dbar_1 \=
\begin{pmatrix}
\delb_A & \cF\\ 0 & \delb
\end{pmatrix}
~, \qquad \Dbar_1^2 \= 0~,
\eeq
where $\cF(\,\cdot\,)= - F_\m\w(\,\cdot\,)^\m$ is the Atiyah map. We need these operators only through order $\ap$.

Define
\beq
\label{eq:KDelta-KZ}
\begin{aligned}
\cK_\D(\D){}^\n{}_\m & \= \nabla^-_\m\D^\n~,\\
\cK_Z(\ccZh){}^\n{}_\m & \= \frac12\nabla^{-\,\n}\ccZh_{\m\bar\rho}\,\dd x^{\bar\rho}~.
\end{aligned}
\eeq
The first row of $\Dbar$ is
\beq
\label{eq:LZ-E-explicit}
\begin{aligned}
\cL_\ccZh(\ccZh) & \= \delb\ccZh - \frac{\ap}{2}\tr\!\left(R^\H\w\cK_Z(\ccZh)\right) + \cO(\ap^2)~,\\
\cE(\aa,\D) & \= - 2\ii\D^\m(\del\wh\o)_\m + \frac{\ap}{2}\tr(F\w\aa) - \frac{\ap}{2}\tr\!\left(R^\H\w\cK_\D(\D)\right) + \cO(\ap^2)~.
\end{aligned}
\eeq
This reproduces the final F-term in \eqref{eq:fterms}. The Bianchi identity gives
$$
\cL_\ccZh^2 \= 0~, \qquad \cL_\ccZh\,\cE + \cE\,\Dbar_1 \= 0~, \qquad \Dbar_1^2 \= 0~,
$$
and hence $\Dbar^2=0$ through order $\ap$.

Let $q=(\aa,\D)$ satisfy $\Dbar_1q=0$. Then $\cE(q)$ is $\cL_\ccZh$-closed and defines an obstruction class
$$
[\cE(q)]\in H_{\cL_\ccZh}^{0,2}\left(T^{*(1,0)}X\right)~.
$$
A $\Dbar$-closed lift of $q$ exists when this obstruction vanishes. The hermitian component of the lift then solves
\beq
\label{eq:lift}
\cL_\ccZh\ccZh \= \! - \cE(\aa,\D)~.
\eeq
Unless the source vanishes, the lift has a non-zero hermitian component. Any two solutions differ by $e$ satisfying $\cL_\ccZh e=0$. If $e$ is $\cL_\ccZh$--exact, it changes only the representative. Otherwise it adds an independent hermitian class.

For example, let $\D$ satisfy $\delb\D=0$. Its lift to $Q_1$ requires a bundle deformation $\aa$ satisfying
$$
\delb_A\aa - F_\m\w\D^\m \= 0~.
$$
The lift to $\cQ$ is then found from \eqref{eq:lift}:
$$
\D\quad\rightsquigarrow\quad(\aa,\D)\quad\rightsquigarrow\quad(\ccZh,\aa,\D)~.
$$

As a second example, set $\D=0$ and take $\delb_A\aa=0$. This bundle deformation lifts to $\cQ$ if $[\cE(\aa,0)]=0$. If $\ccZh_\aa^{\rm p}$ is one solution of
$$
\cL_\ccZh\ccZh_\aa^{\rm p} \= - \cE(\aa,0)~,
$$
the general F-term lift is
\beq
\label{eq:bundleLift}
(\ccZh_\aa^{\rm p} + e,\aa,0)~, \qquad \cL_\ccZh e \= 0~.
\eeq
At leading order, $\cL_\ccZh=\delb$, so $\delb e=0$.

These lifts solve the F-terms but are not yet physical representatives. For the bundle example, the first two D-terms are
\beq
\label{eq:bundleD}
\begin{split}
\delb_A^\dagger\aa + \frac{1}{2}F^{\m\nb}(\ccZh_\aa^{\rm p} + e)_{\m\nb}& \= 0~,\\
\delb^\dagger(\ccZh_\aa^{\rm p} + e)& \= 0~.
\end{split}
\eeq
The third D-term in \eqref{eq:dterms} must also hold. For a fixed total class, the D-terms fix the remaining exact freedom. 

\subsection{Hermitian directions}
\label{s:extension}

We now consider directions that reduce to hermitian moduli at $\ap=0$. We work on a smooth patch with constant cohomology dimensions and no infinitesimal automorphisms.

At $\ap=0$, the hermitian classes lie in
$$
H_\delb(T^{*(1,0)}X)\simeq H^{1,1}(X)~.
$$
We call an $\ap$-corrected direction hermitian if it approaches one of these classes as $\ap\to0$.  In this limit, its $Q_1$ component is cohomologically trivial:
$$
[(\aa_\rho,\D_\rho)] \= 0~,\qquad\text{in}\qquad H_{\Dbar_1}(Q_1)~.
$$
Choose representatives that solve the D-term and F-terms:
\beq
\label{eq:HermReps}
E_\rho \= (e_\rho,\aa_\rho,\D_\rho)~, \qquad \rho \= 1,\ldots,h^{1,1}~.
\eeq
Here $e_\rho$ is the hermitian component. Although $(\aa_\rho,\D_\rho)$ is trivial in $H_{\Dbar_1}^{0,1}(Q_1)$ as $\ap\to0$, the fields $\aa_\rho$ and $\D_\rho$ may still be non-zero because the F- and D-terms are coupled.

Let
$$
\cH_{\rm herm} \= \operatorname{span}\{E_\rho\}
$$
and extend $\{E_\rho\}$ to a local frame $\{E_\rho,\ccY_i\}$ of the physical tangent space, where
\beq
\label{eq:PhysReps}
\ccY_i \= (\wh\ccZ_i,\aa_i,\D_i)~.
\eeq
Projection along the hermitian directions must use the full representatives $E_\rho$, not $(e_\rho,0,0)$. Their compensating components can give non-zero mixed metric coefficients even though \eqref{eq:metric-final-ap2} contains no explicit $\D\wh\ccZ$ pairing.

The same point applies at the standard embedding, where
$$
H^{1,1}(X)\oplus H^{0,1}({\rm End}_0\,TX)\oplus H^{2,1}(X)
$$
is the decomposition of infinitesimal classes \citeDecoupling. This decomposition of classes does not by itself give an orthogonal decomposition of physical representatives. The F- and D-terms can still couple the three components.

The preceding argument is linearised, and throughout this section we assume that we are working at a smooth point of the physical moduli space and that the infinitesimal classes under consideration are unobstructed, so that they are tangent to genuine families of solutions. In complementary mathematical work Wu \cite{Wu:2026pzl} uses the implicit function theorem to construct local families of solutions near a \K point with fixed complex structure, parametrised by an $\ap$-corrected Aeppli class and when bundle deformations are unobstructed, by $H^1(X,\EndE)$, without introducing an independent auxiliary tangent-bundle connection. This gives a nonlinear existence theorem for a related moduli problem and provides a mathematical analogue of the local smoothness assumed here, whereas our analysis treats the coupled physical deformations $\ccY=(\ccZh,\aa,\D)$ and determines their string-derived metric through $\ap^2$. 

\subsection{Pointwise orthogonal projection}
\label{s:orthogonal}

The frame $\{E_\rho,\ccY_i\}$ need not be orthogonal. At a fixed point of moduli space, write the metric as
\beq
\label{eq:metricParts}
h_{\rho\bar\sigma} \= g^\sh(E_\rho,\ol{E}_\sigma)~, \qquad k_{i\bar\sigma} \= g^\sh(\ccY_i,\ol{E}_\sigma)~, \qquad m_{i\jb} \= g^\sh(\ccY_i,\ol{\ccY}_j)~.
\eeq
Let $h^{\rho\bar\sigma}$ denote the inverse of $h_{\rho\bar\sigma}$. All quantities, including $h^{ - 1}$, are expanded through order $\ap^2$.

Project $\ccY_i$ orthogonally to the hermitian directions:
\beq
\label{eq:orthRep}
\ccY_i^\perp \= \ccY_i - k_{i\bar\sigma}h^{\rho\bar\sigma}E_\rho~.
\eeq
Since $E_\rho=(e_\rho,\aa_\rho,\D_\rho)$ is a full physical representative, this subtraction can change all three components of $\ccY_i$.

In the new frame the mixed block vanishes:
\beq
\label{eq:orthMetric}
g^\sh(E_\rho,\ol{E}_\sigma) \= h_{\rho\bar\sigma}~, \qquad g^\sh(\ccY_i^\perp,\ol{E}_\sigma) \= 0~, \qquad g^\sh(\ccY_i^\perp,\ol{\ccY}_j{}^\perp) \= m^\perp_{i\jb}~.
\eeq
The remaining block is
\beq
\label{eq:Schur}
m^\perp_{i\jb} \= m_{i\jb} - k_{i\bar\sigma}h^{\rho\bar\sigma}k_{\rho\jb}~, \qquad k_{\rho\jb} \= g^\sh(E_\rho,\ol{\ccY}_j)~.
\eeq
Thus $m^\perp$ is the Schur complement of the hermitian block. The metric is block diagonal, but neither block need itself be diagonal.

This is a pointwise change of frame, not necessarily a change of moduli coordinates. A coordinate decomposition would require the hermitian directions and their orthogonal complement to be integrable as the moduli vary. We do not assume this.

\section{Conclusion}
\label{s:conclusion}

We have computed the heterotic moduli-space metric through order $\ap^2$ by dimensional reduction in the original Bergshoeff--de Roo variables and then rewriting the result in the hatted variables natural for supersymmetry. In this basis the \K potential retains its familiar form, and its Hessian agrees with the dimensional-reduction metric.

The explicit $\ap^2$ correction arises from the order-$\ap$ deformation of the composite Hull connection. 

The double extension, D-terms and string metric have distinct roles: they organise the F-term constraints, select physical representatives and supply their inner product, respectively. Once representatives have been selected, pointwise projection along the full physical hermitian directions gives the Schur complement of the hermitian block. This is a pointwise change of frame; no coordinate or global product decomposition is assumed. It would be useful to complete the double-extension description through order $\ap^2$ and to compute the Riemann curvature of the resulting moduli-space metric.

\vskip0.5cm
\paragraph*{Acknowledgements.}
We would like to thank S. Picard, R. Plesser, E. Svanes and P. Wu for enlightening conversations and very helpful comments. JM is supported in part by ARC Discovery Project Grants DP240101409 and DP250101828.

\newpage
\appendix

\section{Details of the \texorpdfstring{$\ap^2$}{alpha-prime-squared} moduli-metric calculation}
\label{s:HullDetails}

The Kaluza--Klein reduction and Hull-connection norm below use the original Bergshoeff--de Roo variables. We pass to the hatted variables only in \sref{s:hattedvsunhatted}.

\subsection{Kaluza--Klein reduction}

We reduce on $\IR^{3,1}\times X$. The metric and $B$-field are block diagonal:
\beq
\dd s^2 \= g_{ef}(\cX)\dd\cX^e\dd\cX^f + g_{mn}\bigl(x;y(\cX)\bigr)\dd x^m\dd x^n~, \qquad g_{em} \= B_{em} \= 0~.
\eeq
Four-dimensional gauge vectors are set to zero to preserve Poincar\'e invariance. Write the family of internal gauge connections as
\beq
A_X \= A_m\bigl(x;y(\cX)\bigr)\dd x^m~, \qquad \fD_aA \= \partial_aA - \dd_A\Lambda_a~,
\eeq
Promoting the moduli to scalar fields and including the parameter-space compensators gives
\beq
F_{em} \= (\del_e y^a)(\fD_aA)_m~.
\eeq
The Yang--Mills term contributes
\beq
- \frac{\ap}{4V}\int_X \tr\!\left(\aa_\a\star\ol\aa_\bb\right)~.
\eeq

The tangent bundle and Lorentz group reduce as
\beq\label{eq:spinDecomp}
T\bigl(\IR^{3,1}\times X\bigr) \= T\IR^{3,1}\oplus TX~, \qquad \mathrm{Spin}(1,9)\longrightarrow\mathrm{Spin}(1,3)\times\mathrm{Spin}(6)~.
\eeq
The parameters label a family of compactification backgrounds preserving \eqref{eq:spinDecomp}. For each member of this family, the background Hull connection is block diagonal:
\beq
\Theta^\H_{\mathrm{KK}} \= \Theta^{(4)}\oplus\Theta_X^\H~,
\eeq
where
\beq
\Theta_X^\H(y) \= \Theta^{\LC}\!\left(g_X(y)\right) + \frac12H_X(y)~, \qquad H_X \= \dd B_X + \frac{\ap}{4}\CS(A_X) - \frac{\ap}{4}\CS(\Theta_X^\H)~.
\eeq
$\Theta_X^\H$ is composite, with $\fD_a\Theta_X^\H=\partial_a\Theta_X^\H - \dd_{\Theta_X^\H}\Lambda_a^\Theta$. The covariant variation $\fD_a\Theta_X^\H$ describes a small parameter variation within this family. It is an internal one-form with both Lorentz indices on $TX$. Promoting the parameters to spacetime fields, the spirit of Kaluza--Klein reduction, gives
\beq
\begin{aligned}
&R^\H_{em}{}^{np}& \= (\del_e y^a)\bigl(\fD_a\Theta_X^\H\bigr)_m{}^{np}~,\qquad H_{emn} \= (\del_e y^a)\cB_{a\,mn}~.
\end{aligned}
\eeq
Here $\cB_a$ is the gauge-invariant $B$-field deformation. The spacetime index $e$ in $R^\H_{em}{}^{np}$ is a form index; the Lorentz indices $n,p$ remain internal. The curvature of $\Theta^\H_{\mathrm{KK}}$ has  $R^\H_{mn}{}^{ep}=0$.
 
\subsection{Hull-connection norm}
\label{s:Hullconnnorm}
The curvature-squared term requires
\beq
\frac{\ap}{4V}\int_X \tr\!\left(\fD_\a\Theta_X^\H\star\fD_\bb\Theta_X^\H\right)~.
\eeq
Since this has a factor of $\ap$, we need $\fD\Theta^\H$ through order $\ap$. We drop the subscript $X$ and suppress the parameter indices until the final result. We use the constant-dilaton identities in Appendix~\ref{s:identities}.

Choose the frame $\d e_{ap}=E_a{}^qM_{qp}$, with $M_{pq}=\half(\d g_{pq} + \cB_{pq})$ as in \eqref{eq:Mdef-real}. Varying $\G^\H_m{}^r{}_q=\G^\LC_m{}^r{}_q + \half g^{rs}H_{msq}$ gives
$$
\begin{aligned}
g_{pr}\d\G^\H_m{}^r{}_q \= {}&\nabla_mM_{pq} + \nabla_qM_{mp} - \nabla_pM_{mq}\\[3pt]
& - \half\d g_{pr}H_m{}^r{}_q + \half(\d H - \dd\cB)_{mpq}~.
\end{aligned}
$$
The $\d g\,H$ term comes from varying the inverse metric. We write $\nabla^-$ for the Bismut connection, with $\nabla^-_m v_p=\nabla_m v_p - \half H_{mp}{}^qv_q$ to the order used here. Varying the frame relation \eqref{eq:ThetaGamma} gives
\beq
\label{eq:A2-spin-variation}
\begin{aligned}
\d\Theta^\H_{mab}& \= E_a{}^pE_b{}^q\Big(g_{pr}\d\G^\H_m{}^r{}_q - \nabla^\H_mM_{pq}\Big)\\[3pt]
& \= E_a{}^pE_b{}^q\Big[\nabla^-_qM_{mp} - \nabla^-_pM_{mq} - H_{pq}{}^rM_{mr}\\[3pt]
&\qquad\qquad + \half(\d H - \dd\cB)_{mpq}\Big] + \cO(\ap^2)~.
\end{aligned}
\eeq
Here $\nabla^-$ acts on both indices of $M_{mp}$, including its one-form index $m$ and  the $\nabla_mM_{pq}$ terms cancel. Using $\d g_{pq}=M_{pq} + M_{qp}$, the remaining $H$ terms combine with the conversion to $\nabla^-$.

For a holomorphic parameter variation, \eqref{eq:Mcpx} gives
\beq
\label{eq:A2-M-components}
M_{\tb\nb} \= \D_{\tb\nb}~, \qquad M_{\tb\m} \= \! - \half\ccZ_{\m\tb}~, \qquad M_{\m p} \= 0~.
\eeq
Since Bismut preserves type, these components can be substituted directly into \eqref{eq:A2-spin-variation}. By \eqref{eq:h-deformation}, the term $\half(\d H - \dd\cB)_{\tb\nb\m}$ gives the Chern--Simons contributions. Thus
\beq
\label{eq:DthccPC}
\fD\Theta^\H_{\tb\nb\m} \= \ccP_{\m\nb\tb} + \ap C^F_{\m\nb\tb} + \ap C^R_{\m\nb\tb} + \cO(\ap^2)~,
\eeq
where
\beq
\label{eq:Pdef}
\begin{aligned}
\ccP_{\m\nb\tb} \= {}&\nabla^-_\m\D_{\tb\nb} + \half\nabla^-_\nb\ccZ_{\m\tb} + H_{\m\nb}{}^{\rb}\D_{\tb\rb} - \half H_{\m\nb}{}^\r\ccZ_{\r\tb}~.
\end{aligned}
\eeq
The Chern--Simons terms are evaluated to zeroth order in $\ap$:
$$
\begin{aligned}
C^F_{\m\nb\tb}& \= \! - \qrt\tr\!\left(\aa_{\tb}F_{\m\nb} - \aa_{\nb}F_{\m\tb}\right)~,\\[3pt]
C^R_{\m\nb\tb}& \= \half\left(R^\H{}_{\m\nb}{}^{\s\lb}\ccP_{\s\lb\tb} - R^\H{}_{\m\tb}{}^{\s\lb}\ccP_{\s\lb\nb}\right)~.
\end{aligned}
$$
For future reference, the pure components are
\beq
\begin{aligned}
\fD\Theta^{\H\,0,1}_{\m\n}& \= \! - \half\left(\nabla^-_\m\ccZ_\n - \nabla^-_\n\ccZ_\m + H_{\m\n}{}^\r\ccZ_\r\right) + \cO(\ap^2)~,\\[5pt]
\fD\Theta^{\H\,0,1}_{\mb\nb}& \= \! - \nabla^-_\mb\D_\nb + \nabla^-_\nb\D_\mb - H_{\mb\nb}{}^{\rb}\D_\rb + \cO(\ap^2)~.
\end{aligned}
\eeq
Here we use right contraction as in \eqref{eq:identities-right-contraction}.

For the norm, we use the Chern connection $\nabla^\CH$ on form indices and Bismut on the lowered vector index of $\D$. Denote this tensor-product derivative by $\nabla^{\CH,-}$. The first index of $\D_{\nb\tb}$ is the form index and the second is the lowered vector index, so
\beq
\label{eq:A2-product-derivative}
\nabla^{\CH,-}_\m\D_{\nb\tb} \= \nabla^\CH_\m\D_{\nb\tb} - H_{\m\tb}{}^{\rb}\D_{\nb\rb} + \cO(\ap^2)~.
\eeq
Chern also acts on derivative indices. Both connections preserve type and are metric compatible. On the $(1,1)$-component of $\ccZ$, the Hull derivative agrees with Chern:
\beq
\label{eq:A2-Hull-Z-derivative}
\nabla^\H_\nb\ccZ_{\m\tb} \= \nabla^\CH_\nb\ccZ_{\m\tb} + \cO(\ap^2)~.
\eeq
To zeroth order in $\ap$, $\D_{\mb\nb}$ is symmetric and the representatives obey the K\"ahler F-terms and D-terms in \eqref{eq:susyeom}. Symmetrising \eqref{eq:Pdef} gives
\beq
\label{eq:A2-P-symmetry}
\ccP_{\m(\nb\tb)} \= \half\left(\nabla^{\CH,-}_\m\D_{\nb\tb} + \nabla^{\CH,-}_\m\D_{\tb\nb}\right) 
- \half\nabla^\CH_{(\nb}\ccZ_{\tb)\m} + \cO(\ap^2)~.
\eeq
The antisymmetric part of $\ccP$ is $\cO(\ap)$. Since $C^F$ and $C^R$ are antisymmetric in $\nb,\tb$, their pairing in \eqref{eq:DthccPC} first enters the metric at order $\ap^3$. The pure tangent components also vanish to zeroth order by the K\"ahler F-terms. For the $(1,0)$ form components, $M_{\m p}=0$ and type preservation in \eqref{eq:A2-spin-variation} give $\fD\Theta^\H_{\m pq}=\cO(\ap)$. Their squared norms first enter at order $\ap^3$.

Define
\beq
\label{eq:IccP}
I_\ccP \= \int_X\vol\,\ccP_{\m\nb\tb}\ccPb^{\m\nb\tb}~.
\eeq
Then
\beq
\label{eq:A2-norm-reduction}
\frac{\ap}{4V}\int_X\tr\!\left(\fD\Theta^\H\star\fDb\Theta^\H\right) \= \! - \frac{\ap}{2V}I_\ccP + \cO(\ap^3)~.
\eeq
The symmetric and antisymmetric parts in \eqref{eq:A2-P-symmetry} are orthogonal. The antisymmetric derivatives are $\cO(\ap)$, so
\beq
\label{eq:Pleading}
\begin{aligned}
I_\ccP& \= I_\D + \qrt I_\ccZ + I_{\D\ccZb} + I_{\ccZ\Db} + \cO(\ap^2)~,\\[3pt]
I_\D& \= \int_X\vol\,\nabla^{\CH,-}_\m\D_{\tb\nb}\,\big(\nabla^{\CH,-}\big)^\m\ol\D^{\tb\nb}~,\\[3pt]
I_\ccZ& \= \int_X\vol\,\nabla^\CH_\nb\ccZ_{\m\tb}\,\big(\nabla^\CH\big)^\nb\ccZb^{\m\tb}~,\\[3pt]
I_{\D\ccZb}& \= \half\int_X\vol\,\nabla^{\CH,-}_\m\D_{\tb\nb}\,\big(\nabla^\CH\big)^\nb\ccZb^{\m\tb}~.
\end{aligned}
\eeq
The remaining mixed term is obtained by complex conjugation.

We first record the identities needed for integration by parts. The torsion of the Chern connection is $\ap^2$ by the balanced equation; hence, the integral of a total divergence with respect to Chern  integrates to $\cO(\ap^2)$ on compact $X$. The same holds for contractions formed with $\nabla^{\CH,-}$, since the connection on the vector index is metric compatible. Equations \eqref{eq:susyeom}  imply
$$
\begin{aligned}
\nabla^{\CH,-}_{[\m}\ol\D_{\r]\s}& \= \cO(\ap)~, &\big(\nabla^{\CH,-}\big)^\m\ol\D_{\m\s}& \= \cO(\ap)~,\\[3pt]
\nabla^\CH_{[\l}\ccZb_{\r]\sb}& \= \cO(\ap)~, &\big(\nabla^\CH\big)^\l\ccZb_{\l\sb}& \= \cO(\ap)~,
\end{aligned}
$$
and their conjugates. Differentiated F-term and D-term contributions below are integrated by parts, giving products of two $\cO(\ap)$ terms.

The balanced condition and the $H$ equation give $g^{\m\lb}(\dd H)_{\m\r\lb\nb}=2\nabla^\k H_{\k\r\nb} + \cO(\ap^2)$. Equations \eqref{eq:identities-gravitonEOM} and \eqref{eq:traceDH} therefore imply
\beq
\label{eq:A2-H-divergence}
\nabla^\k H_{\k\r\nb} \= \! - \frac{\ap}{4}\cE^\H_{\r\nb} + \cO(\ap^2) \= \Ric_{\r\nb} + \cO(\ap^2)~.
\eeq
The Hull and Chern connection symbols $\G_m{}^\r{}_\s$ agree through order $\ap$. The remaining curvature difference is quadratic in $H$, so
$$
R^\CH_{mn}{}^\r{}_\s \= R^\H_{mn}{}^\r{}_\s + \cO(\ap^2)~.
$$
The terms linear in $\nabla H$ cancel between Hull and Bismut. Using the preceding Chern--Hull relation gives
$$
R^-_{mn}{}^\r{}_\s + R^\CH_{mn}{}^\r{}_\s \= 2R_{mn}{}^\r{}_\s + \cO(\ap^2)~.
$$
 Using \eqref{eq:Hull-curvature-trace}, \eqref{eq:traceDH}, \eqref{eq:trRiemannap2} and \eqref{eq:RictrR}, we obtain
\beq
\label{eq:A2-Hermitian-traces}
\begin{aligned}
g^{\m\lb}R^\CH_{\m\lb}{}^\k{}_\r& \= \Ric_\r{}^\k + \cO(\ap^2)~,\qquad g^{\m\lb}R^-_{\m\lb}{}^\k{}_\r\= \! - \Ric_\r{}^\k + \cO(\ap^2)~,\\[6pt]
g^{\m\lb}R^\CH_{\r\lb}{}^\k{}_\m& \= \half\Ric_\r{}^\k + \half\nabla^\m H_{\r\m}{}^\k + \cO(\ap^2) \= \cO(\ap^2)~.
\end{aligned}
\eeq
In the last line, \eqref{eq:A2-H-divergence} gives $\nabla^\m H_{\r\m}{}^\k= - \Ric_\r{}^\k + \cO(\ap^2)$.

We first evaluate $I_\D$. Integrating by parts gives
\beq
\label{eq:A2-D-first-IBP}
I_\D\cong - \int_X\vol\,\D^{\r\s}g^{\m\lb}\nabla^{\CH,-}_\m\nabla^{\CH,-}_\lb\ol\D_{\r\s} + \cO(\ap^2)~.
\eeq
Commuting the derivatives produces Chern curvature on $\r$ and Bismut curvature on $\s$. Their traces in \eqref{eq:A2-Hermitian-traces} cancel in the integral:
$$
\D^{\r\s}\Big(\Ric_\r{}^\k\ol\D_{\k\s} - \Ric_\s{}^\k\ol\D_{\r\k}\Big) \= \cO(\ap^2)~.
$$
We then exchange $\m,\r$ using the F-term and commute once more to apply the D-term. Chern's torsion has no mixed components, so
\beq
\begin{aligned}
I_\D&\cong - \int_X\vol\,\D^{\r\s}g^{\m\lb}\nabla^{\CH,-}_\lb\nabla^{\CH,-}_\r\ol\D_{\m\s}\\[3pt]
&\cong - \int_X\vol\,\D^{\r\s}g^{\m\lb}\Big(R^\CH_{\r\lb}{}^\k{}_\m\ol\D_{\k\s} + R^-_{\r\lb}{}^\k{}_\s\ol\D_{\m\k}\Big) + \cO(\ap^2)~.
\end{aligned}
\eeq
The first curvature term vanishes by the last line of \eqref{eq:A2-Hermitian-traces}. Hence
\beq
\label{eq:IDinter}
\begin{aligned}
I_\D&\cong\int_X\vol\,\D_{\mb\nb}\ol\D_{\s\r}R^{-\,\s\mb\r\nb}\\[3pt]
& \= \int_X\vol\,\D_{\mb\nb}\ol\D_{\s\r}R^{\H\,\s\mb\r\nb} + \cO(\ap^2)~.
\end{aligned}
\eeq
For the last equality, expand the two curvatures about Levi--Civita. Each $\nabla H$ term contracts an antisymmetric pair with $\D$ or $\ol\D$, and is $\cO(\ap^2)$ since $\D_{[\mb\nb]}=\cO(\ap)$.

We next evaluate $I_\ccZ$. Integrating by parts, exchanging $\l,\r$ with the F-term and applying the D-term gives
\beq
\label{eq:A2-Z-first-IBP}
\begin{aligned}
I_\ccZ&\cong - \int_X\vol\,\ccZ^{\r\sb}g^{\l\nb}\nabla^\CH_\nb\nabla^\CH_\l\ccZb_{\r\sb}\\[3pt]
&\cong - \int_X\vol\,\ccZ^{\r\sb}g^{\l\nb}\big[\nabla^\CH_\nb,\nabla^\CH_\r\big]\ccZb_{\l\sb}\\[3pt]
& \= - \int_X\vol\,\ccZ^{\r\sb}g^{\l\nb}\Big(R^\CH_{\r\nb}{}^\k{}_\l\ccZb_{\k\sb} + R^\CH_{\r\nb}{}^\kb{}_\sb\ccZb_{\l\kb}\Big) + \cO(\ap^2)~.
\end{aligned}
\eeq
Commuting $\nabla^\CH_\nb$ and $\nabla^\CH_\r$ gives no torsion term, since $T^\CH_{\nb\r}{\,}^p=0$. The first curvature term again vanishes by \eqref{eq:A2-Hermitian-traces}; the second gives
\beq
\label{eq:A2-Hull-Z-pairing}
I_\ccZ \= \int_X\vol\,\ccZ_{\r\mb}\ccZb_{\s\nb}R^{\H\,\s\mb\r\nb} + \cO(\ap^2)~.
\eeq

It remains to evaluate the mixed terms. Expanding \eqref{eq:A2-product-derivative} and integrating the Chern derivative by parts gives
\beq
\label{eq:DZmix}
\begin{aligned}
I_{\D\ccZb}& \= \half\int_X\vol\,\Big(\nabla^\CH_\m\D_{\tb\nb} - H_{\m\nb}{}^\rb\D_{\tb\rb}\Big)\big(\nabla^\CH\big)^\nb\ccZb^{\m\tb} + \cO(\ap^2)\\[3pt]
&\cong - \half\int_X\vol\,\D^{\l\r}\Big(\nabla^\CH_\m\nabla^\CH_\r\ccZb^\m{}_\l - H_{\m\r}{}^\k\nabla^\CH_\k\ccZb^\m{}_\l\Big) + \cO(\ap^2)~.
\end{aligned}
\eeq
The Chern curvature has no $(2,0)$ component, while its torsion is $T^\CH_{\m\r}{}^\k= - H_{\m\r}{}^\k + \cO(\ap^2)$. Therefore
$$
\big[\nabla^\CH_\m,\nabla^\CH_\r\big]\ccZb^\m{}_\l \= H_{\m\r}{}^\k\nabla^\CH_\k\ccZb^\m{}_\l + \cO(\ap^2)~,
$$
which cancels the explicit $H$ term in \eqref{eq:DZmix}. A second integration by parts gives
\beq
\label{eq:IDccZ}
\begin{aligned}
I_{\D\ccZb}&\cong\half\int_X\vol\,\big(\nabla^\CH_\r\D^{\l\r}\big)\big(\nabla^\CH_\m\ccZb^\m{}_\l\big) \= \cO(\ap^2)~,\\[3pt]
I_{\ccZ\Db}& \= \cO(\ap^2)~.
\end{aligned}
\eeq
The first divergence is $\cO(\ap)$ by the leading symmetry of $\D$ and its D-term. For the second, the leading $(1,1)$-form is K\"ahler harmonic and has constant trace, so $\nabla^\CH_\m\ccZb^\m{}_\l=\cO(\ap)$.

Substituting \eqref{eq:IDinter}, \eqref{eq:A2-Hull-Z-pairing} and \eqref{eq:IDccZ} into \eqref{eq:A2-norm-reduction} and \eqref{eq:Pleading} gives
\beq
\label{eq:Hull-norm-IP}
\begin{aligned}
&\frac{\ap}{4V}\int_X\tr\!\left(\fD\Theta^\H\star\fD\Theta^\H\right)\\[3pt]
&\qquad\cong - \frac{\ap}{2V}\int_X\vol\,\left(\D_{\mb\nb}\,\Db_{\s\r} + \qrt\ccZ_{\r\mb}\,\ccZb_{\s\nb}\right)R^{\H\,\s\mb\r\nb} + \cO(\ap^3)~.
\end{aligned}
\eeq

\subsection{The metric in hatted and unhatted variables}
\label{s:hattedvsunhatted}
The hatted fields and hermitian deformation are defined in \eqref{eq:hatted-fields} and \eqref{eq:hatted-Z}. In the following comparison, $\ccZ$ and $\wh\ccZ$ denote their $(1,1)$ components. We first compare the unsuppressed $L^2$ pairings in the two field bases. Keeping the Hodge star fixed at its unhatted value, the field redefinition of the hermitian deformation gives
\beq
\label{eq:Zhpairing}
\int_X \wh\ccZ\star\wh\ccZb \cong \int_X\ccZ\star\ccZb - \frac{\ap^2}{4}\int_X\left(\d\cE^\H\star\ccZb - \ccZ\star\ol\d\cE^\H\right) + \cO(\ap^3)~.
\eeq
The cross term can be reduced using the leading-order harmonicity of $\ccZb$ and \eqref{eq:traceDH-inverse}:
\beq
\label{eq:hatted-E-pairing}
\begin{split}
\ap\int_X\d\cE^\H\w\star\ccZb &\cong\int_X\vol\, \ccZ{}^{\m\lb}\ccZb{}^{\r\nb} (\dd H)_{\m\r\lb\nb}\\
&\cong - \frac{\ap}{2}\int_X\vol\,\Big[ \ccZ{}^{\r\sb}\ccZb_{\tau\sb} + \ccZb{}^{\r\sb}\ccZ_{\tau\sb} \Big]\cE^\H_\r{}^\tau~.
\end{split}
\eeq
The trace $c=g^{\m\nb}\ccZ_{\m\nb}$ is constant and $\o\w\ccZb$ is harmonic at leading order. The trace contribution omitted in the second line is beyond the required order, since
$$
\langle\L(\dd H),\ccZb\rangle \= \langle\dd H,\o\w\ccZb\rangle
$$
and
\beq
- \frac{\ap}{2}c\int_X\vol\,\ccZb{}^{\m\nb}\cE^\H_{\m\nb} \sim c\langle\dd H,\o\w\ccZb\rangle \= \cO(\ap^2)~.
\eeq
Thus the field variation in \eqref{eq:Zhpairing} is
\beq
\label{eq:Zhpairing_refined}
\int_X\wh\ccZ\star\wh\ccZb \cong\int_X\ccZ\star\ccZb + \frac{\ap^2}{4}\int_X\vol\,\Big[ \ccZ{}^{\r\sb}\ccZb_{\tau\sb} + \ccZ_{\tau\sb}\ccZb{}^{\r\sb} \Big]\cE^\H_\r{}^\tau ~.
\eeq
The change of Hodge star produces the opposite correction:
\beq
\label{eq:Zpairing_whstar}
\begin{aligned}
\int_X\ccZ\;\wh\star\;\ccZb &\cong\int_X\ccZ\star\ccZb - \frac{\ap^2}{4}\int_X\vol\,\Bigl( \ccZ_{\m\nb}\ccZb_{\tb\r}\cE^{\H\,\tb\m}g^{\r\nb} + \ccZ_{\m\nb}\ccZb_{\tb\r}g^{\m\tb}\cE^{\H\,\nb\r} \Bigr)\\[5pt]
& \= \int_X\ccZ\star\ccZb - \frac{\ap^2}{4}\int_X\vol\,\Bigl( \ccZ{}^{\r\sb}\ccZb_{\t\sb} + \ccZ_{\t\sb}\ccZb{}^{\r\sb} \Bigr)\cE^\H_\r{}^\t~.
\end{aligned}
\eeq
Combining \eqref{eq:Zhpairing_refined} and \eqref{eq:Zpairing_whstar} makes the cancellation between the field redefinition and the Hodge-star variation explicit:
\beq
\label{eq:Zhpairing_refined_final}
\int_X\wh\ccZ\;\wh\star\;\wh\ccZb \cong\int_X\ccZ\star\ccZb~.
\eeq

The complex-structure pairing behaves similarly:
\beq
\label{eq:DDbhat}
\begin{split}
\int_X\,\D{}^\m\;\wh\star\;\Db{}^\nb \;\wh g_{\m\nb} &~\cong~\int_X\volh\;\D_{\mb}{}^\n\;\Db_{\r}{}^\sb \;\wh g^{\r\mb} \;\wh g_{\n\sb}\\
&~\cong~\int_X\D{}^\m\star\Db{}^\nb g_{\m\nb} - \frac{\ap^2}{4}\int_X\vol\,\Bigl( \D_{\mb}{}^\n\Db_{\r}{}^\sb\cE^{\H\,\mb\r}g_{\n\sb} - \D_{\mb}{}^\n\Db_{\r}{}^\sb g^{\mb\r}\cE^\H_{\n\sb} \Bigr)\\
&\cong\int_X\D{}^\m\star\Db{}^\nb g_{\m\nb}~.
\end{split}
\raisetag{1cm}
\eeq
In the last equality we used $\D_{[\mb\nb]}=\cO(\ap)$. Together with $\wh V=V + \cO(\ap^3)$ \citeMP, equations \eqref{eq:Zhpairing_refined_final} and \eqref{eq:DDbhat} show that the normalized hatted and unhatted pairings agree on--shell through order $\ap^2$.

\subsection{The \K potential}
\label{s:KahlerPot}
The \K potential is
\beq
K \= K_1 + K_2 \= \! - \log\left(\frac{4}{3}\int_X(\wh\o)^3\right) - \log\left(\ii\int_X\Omega\w\Ob\right) + \cO(\ap^3)~.
\eeq
For $K_1$,
\beq\notag
\begin{aligned}
\del_\a\del_\bb K_1 &= - \del_\a\left(\frac{1}{\int_X(\wh\o)^3}\int_X3(\del_\bb\wh\o)(\wh\o)^2\right)\\
&= - \frac{1}{\wh V}\int_X\fD_\a\wh\o^{0,2}\;\wh\star\;\fD_\bb\wh\o^{2,0} + \frac{1}{\wh V}\int_X\fD_\a\wh\o^{1,1}\;\wh\star\;\fD_\bb\wh\o^{1,1} - \frac{1}{2\wh V}\int_X(\wh\o)^2\del_\a\del_\bb\wh\o~.
\end{aligned}
\eeq
Using $(\d\wh\o)^{p,q}\cong\d y^\a\fD_\a\wh\o^{p,q}$ and the hatted F-terms and Bianchi identity \citeMP,
\beq\notag
\begin{aligned}
- \frac{1}{2\wh V}\int_X(\wh\o)^2\del_\a\del_\bb\wh\o &=\frac{\ii}{4\wh V}\int_X\ii(\wh\o)^2\{\del_\a,\del_\bb\}\wh\o^{1,1}\\
&= - \frac{\ii\ap}{8\wh V}\int_X(\wh\o)^2\left[\tr\left(\fD_\a A\,\fD_\bb A\right) - \tr\left(\fD_\a\Theta^\H\,\fD_\bb\Theta^\H\right)\right] + \frac{2}{\wh V}\int_X\fD_\a\wh\o^{0,2}\;\wh\star\;\fD_\bb\wh\o^{2,0}~.
\end{aligned}
\eeq
In constant-dilaton gauge,
$$
- \frac{1}{2\wh V}\D_\a{}^\r(\fD_\bb\wh\o^{2,0})_\r(\wh\o)^2 \= \frac{1}{\wh V}\fD_\a\wh\o^{0,2}\;\wh\star\;\fD_\bb\wh\o^{2,0}~,
$$
Together with this identity, the relations $\dd((\wh\o)^2)\cong0$ and $\ap(\wh\o)^2F\cong0$ give
\beq
\label{eq:ddK1}
\del_\a\del_\bb K_1 \= \! - \frac{\ii\ap}{8\wh V}\int_X(\wh\o)^2\left[\tr\left(\fD_\a A\,\fD_\bb A\right) - \tr\left(\fD_\a\Theta^\H\,\fD_\bb\Theta^\H\right)\right] + \frac{1}{\wh V}\int_X\fD_\a\wh\o\;\wh\star\;\fD_\bb\wh\o~.
\eeq
For $K_2$,
\beq\notag
\begin{aligned}
\del_\a\del_\bb K_2 & \= - \del_\a\left(\frac{1}{\int_X\ii\O\w\Ob}\int_X\ii\O\w\del_\bb\Ob\right)\\
& \= \frac{1}{\wh V}\int_X\volh\,\D_{\a(\mb\nb)}\ol\D_\bb{}^{(\mb\nb)} - \frac{1}{\wh V}\int_X\volh\,\D_{\a[\mb\nb]}\ol\D_\bb{}^{[\mb\nb]}~.
\end{aligned}
\eeq
Indices are raised with $\wh g$. Since
$$
\fD_\a\wh\o^{0,2}\;\wh\star\;\fD_\bb\wh\o^{2,0} \= 2\D_{\a[\mb\nb]}\ol\D_\bb{}^{[\mb\nb]}\volh~,
$$
combining with \eqref{eq:ddK1} gives
\beq
\label{eq:moduli-metric_b}
\begin{aligned}
g^{\sh}_{\a\bb} \= \frac{1}{\wh V}\int_X\Bigg(&\D_\a{}^\m\;\wh\star\;\ol\D_\bb{}^\nb\wh g_{\m\nb} + \frac14\wh\ccZ_\a^{1,1}\;\wh\star\;\wh\ccZb_\bb^{1,1} - \frac{\ap}{4}\tr(\aa_\a\star\ol\aa_\bb)\\
&\qquad + \frac{\ap}{4}\tr(\fD_\a\Theta^\H\star\fD_\bb\Theta^\H)\Bigg) + \cO(\ap^3)~.
\end{aligned}
\eeq
Using \eqref{eq:Zhpairing_refined_final}, \eqref{eq:DDbhat} and $\wh V=V + \cO(\ap^3)$, we obtain the same metric in the original variables:
\beq
\label{eq:moduli-metric_c}
\begin{aligned}
g^{\sh}_{\a\bb} \= \frac{1}{V}\int_X\Bigg(&\D_\a{}^\m\star\ol\D_\bb{}^\nb g_{\m\nb} + \frac14\ccZ_\a^{1,1}\star\ccZb_\bb^{1,1} - \frac{\ap}{4}\tr(\aa_\a\star\ol\aa_\bb)\\
&\qquad + \frac{\ap}{4}\tr(\fD_\a\Theta^\H\star\fD_\bb\Theta^\H)\Bigg) + \cO(\ap^3)~.
\end{aligned}
\eeq

\section{Conventions and useful identities}
\label{s:identities}

We write $\nabla$ and $R_{mn}{}^p{}_q$ for the Levi--Civita connection and curvature, with $\Ric_{mn} \= R_{pm}{}^p{}_n$. In complex coordinates
\beq
\label{eq:curvconventions}
[\nabla_m,\nabla_n]v_p \= \! - R_{mn}{}^q{}_{p} v_q~, \qquad [\nabla_m,\nabla_n]w^p \= R_{mn}{}^p{}_{q} w^q~.
\eeq
Under an integral, $\cong$ denotes equality up to integration by parts, F-terms, D-terms and terms that first enter the metric at order $\ap^3$. The identities below hold on the supersymmetric background to the order displayed. Constant-dilaton gauge is used in the Ricci and trace relations.

The Hull connection and curvature are
\beq
\label{eq:Hull-curvature-LC}
\Theta^\H \= \Theta^\LC + \frac12H~, \qquad R^\H_{mnpq} \= R_{mnpq} + \nabla_{[m}H_{n]pq} + \frac12H_{[m|p|}{}^rH_{n]rq}~.
\eeq
Here $H=\cO(\ap)$. On the supersymmetric background, the Bismut connection $\Theta^- =\Theta^\LC - \frac12H$ has $SU(3)$ holonomy, and
\beq
\label{eq:Hull-Bismut-exchange}
R^\H_{mnpq} \= R^-_{pqmn} + \frac12(\dd H)_{mnpq}~.
\eeq
Hence the components with a $(0,2)$ form-index  pair is
\beq
\label{eq:Hull-curvature-02}
R^\H_{\mb\nb pq} \= \frac12(\dd H)_{\mb\nb pq} + \cO(\ap^2)~.
\eeq
The Hermitian trace is
\beq
\label{eq:Hull-curvature-trace}
g^{\m\nb}R^\H_{\m\nb pq} \= \frac12g^{\m\nb}(\dd H)_{\m\nb pq} + \cO(\ap^2)~.
\eeq
For Levi--Civita,
\beq
\label{eq:LC-curvature-20-02}
R_{\m\r\kb\l} \= \nabla_{[\m} H_{\r]\l\kb} + \cO(\ap^2)~.
\eeq
The non-\K index-exchange identity is
\beq
\label{eq:dHtr}
R_{\rho\lb\mu\kb} - R_{\mu\lb\rho\kb} \= \! - 2R_{\mu[\kb\lb]\rho} = - \frac14(\dd H)_{\mu\rho\lb\kb} + \cO(\ap^2)~.
\eeq

For $T=\dd^c\o$, the Levi--Civita symbols are
\beq\label{eq:LCsymbols}
\begin{aligned}
\G_{\m}{}^{\n}{}_{\r} & \= g^{\n\sb}\del_\m g_{\r\sb} - \frac12T_{\m}{}^{\n}{}_{\r}~, \qquad& \G_{\m}{}^{\nb}{}_{\r} & \= 0~,\\[5pt]
\G_{\m}{}^{\n}{}_{\rb} & \= \frac12T_{\m}{}^{\n}{}_{\rb}~, & \G_{\m}{}^{\nb}{}_{\rb} & \= \! - \frac12T_{\m}{}^{\nb}{}_{\rb}~.
\end{aligned}
\eeq
To the order used here, $T$ may be replaced by $H$.

For suppressed form indices we use right contraction. For one-forms $A$ and $B$,
\beq\label{eq:identities-right-contraction}
\begin{aligned}
X_m & \= \frac{1}{(r - 1)!}X_{n_1\cdots n_{r - 1}m}\, \dd x^{n_1}\w\cdots\w\dd x^{n_{r - 1}}\\[5pt]
& \= ( - 1)^{r - 1}\iota_{\partial_m}X~,\\[5pt]
(A\w B)_m& \= A\,B_m - A_mB~.
\end{aligned}
\eeq
Define
\beq
\label{eq:hatted-reduction-E}
h \= \tr|F|^2 - \tr|R^\H|^2~, \qquad \cE^\H_{mn} \= \tr(F_{mp}F_n{}^p) - \tr(R^\H_{mp}R^{\H}{}_n{}^p)~.
\eeq
On this background,
$$
\cE^\H_{\m\nb}=\cO(1)~, \qquad g^{\m\nb}\cE^\H_{\m\nb}=h + \cO(\ap)~, \qquad h=\cO(\ap)~.
$$
The Bianchi identity is
\beq
\dd H \= \frac{\ap}{4}\tr F\w F - \frac{\ap}{4}\tr R^\H\w R^\H~.
\eeq
In constant-dilaton gauge the graviton equation gives
\beq
\label{eq:identities-gravitonEOM}
\Ric_{\r\nb} \= \! - \frac{\ap}{4}\cE^\H_{\r\nb} + \cO(\ap^2)~.
\eeq
The supersymmetry identity gives
\beq
\label{eq:traceDH}
g^{\m\lb}(\dd H)_{\m\r\lb\nb} \= \! - \frac{\ap}{2}\cE^\H_{\r\nb} + \cO(\ap^2)~.
\eeq
In complex dimension three the inverse is
\beq
\label{eq:traceDH-inverse}
\begin{split}
(\dd H)_{\m\r\lb\nb} \= \! - \frac{\ap}{2}\Big[& g_{\m\lb}\cE^\H_{\r\nb} - g_{\r\lb}\cE^\H_{\m\nb} - g_{\m\nb}\cE^\H_{\r\lb} + g_{\r\nb}\cE^\H_{\m\lb}\\
& - \frac12g^{\n\tb}\cE^\H_{\n\tb} \left(g_{\m\lb}g_{\r\nb} - g_{\r\lb}g_{\m\nb}\right) \Big] + \cO(\ap^2)~.
\end{split}
\eeq
The $H$ equation gives
\beq
\label{eq:trace-dH-31}
(\dd H)_{\m\n\r}{}^\r \= 2H^{\r\lb}{}_{[\m}H_{\n]\r\lb} \= \cO(\ap^2)~.
\eeq
The remaining contractions used in Appendix~\ref{s:HullDetails} are
\beq\label{eq:trRiemannap2}
g^{\l\mb} R_{\l\mb\r\nb} \= \! - \Ric_{\r\nb} + \half (\dd H){}^\lb{}_{\r\lb\nb} \= \cO(\ap^2)~,
\eeq
and
\beq\label{eq:RictrR}
\Ric_{\r\tb} \= 2 R^\m{}_{\r\m\tb} + g^{\l\mb} R_{\l\mb\r\tb} \cong 2 R^\m{}_{\r\m\tb} ~.
\eeq

\subsection{Deformation identities}

Through order $\ap$, the Chern component form of the F-terms and D-terms used in Appendix~\ref{s:HullDetails} is
\beq
\label{eq:susyeom}
\begin{gathered}
2\nabla^\CH_{[\mb}\D_{\nb]\tb} \= H_{\mb\nb}{}^\rb\D_{\rb\tb}~, \qquad \big(\nabla^\CH\big)^{\mb}\D_{\mb\nb} \= \cG_{\nb}~,\\[2pt]
\nabla^\CH_\m\ol\D_{\n\r} \= \nabla^\CH_\n\ol\D_{\m\r} + H_{\m\n}{}^\k\Db_{\k\r}~, \qquad \big(\nabla^\CH\big)^\m\ol\D_{\m\r} \= \overline{\cG}_\r~,\\[2pt]
\nabla^\CH_{[\mb}\ccZ_{\rb]\n} \= \! - \half\cE_{\mb\rb\n} + \half H_{\mb\rb}{}^\kb\ccZ_{\kb\n} + \half H_{[\mb|\n|}{}^\k\ccZ_{\rb]\k}~,\\[2pt]
\big(\nabla^\CH\big)^{\nb}\ccZ_{\m\nb} \= \half g^{\l\nb}H_{\l\m}{}^\k\ccZ_{\k\nb}~.
\end{gathered}
\eeq
Here $\cE_{\mb\rb\n}$ and $\cG_{\nb}$ are the order-$\ap$ F-term and D-term sources. They are distinct from $\cE^\H_{mn}$ in \eqref{eq:hatted-reduction-E}. 

Finally, let
$$
M_{a\,mn} \= \frac12\left(\del_a g_{mn} + \cB_{a\,mn}\right)~.
$$
In holomorphic gauge,
\beq
\begin{aligned}
M_{\a\,\mb\nb}& \= \D_{\a\,\mb\nb}~, & M_{\a\,\mb\n}& \= \frac12\ccZ_{\a\,\mb\n}~, & M_{\a\,\m\n}& \= M_{\a\,\m\nb} \= 0~,\\
M_{\bb\,\m\n}& \= \ol\D_{\bb\,\m\n}~, & M_{\bb\,\m\nb}& \= \frac12\ccZb_{\bb\,\m\nb}~, & M_{\bb\,\mb\n}& \= M_{\bb\,\mb\nb} \= 0~.
\end{aligned}
\eeq
Although components of $M$ may vanish, their Levi--Civita derivatives need not:
\beq
\begin{split}
 \nabla_\m M_{\a\,\n\s} & \= 0~, \quad \nabla_\m M_{\a\,\n\sb} \= 0~,\quad \nabla_\m M_{\a\,\nb\s} \= \! - \half \nabla_\m \ccZ_{\a\,\s\nb} ~, \\[5pt]
 \nabla_\m M_{\a\,\nb\rb} & \= \nabla_\m\D_{\a\,\nb\rb} + \qrt H_\m{}^\s{}_\rb \ccZ_{\a\,\s\nb}~, \\[5pt]
  \nabla_\nb M_{\a\,\m\r} & \= \qrt H_\nb{}^\sb{}_\m \ccZ_{\a\,\r\sb}~, \quad \nabla_\nb M_{\a\,\mb\r} \= \! - \half \nabla_\nb \ccZ_{\a\,\r\mb} - \half H_\nb{}^\sb{}_\r \D_{\a\,\mb\sb}~,\\[5pt]
  \nabla_\nb M_{\a\,\mb\rb} & \= \nabla_\nb \D_{\a\,\mb\rb}~,\qquad \nabla_\nb M_{\a\,\m\rb} \= \! - \half H_\nb{}^\sb{}_\m \D_{\a\,\sb\rb}~,
\end{split}
\eeq
where
$$
\nabla_\m \ccZ_{\a\,\nb\s} \= \del_\m \ccZ_{\a\,\nb\s} - \G_\m{}^\rb{}_\nb \ccZ_{\a\,\rb\s} - \G_\m{}^\r{}_\s \ccZ_{\a\,\nb\r}~, \qquad \ccZ_{\a\,\m\n} \= \ccZ_{\a\,\mb\nb} \= 0~.
$$


\providecommand{\href}[2]{#2}\begingroup\raggedright\endgroup

\end{document}